\documentclass[aps,pra,twocolumn,amsmath,amssymb,amsfonts,floatfix,showpacs]{revtex4-1}
\usepackage{graphicx}
\usepackage{bm}
\newcommand{\tst}{\textstyle}

\newcommand{\mbf}{\mathbf}
\newcommand{\mrm}{\mathrm}

\begin{document}

\author{Zbigniew Idziaszek}
\affiliation{Institute of Theoretical Physics, University of Warsaw,
00-681 Warsaw, Poland}
\author{Tommaso Calarco}
\affiliation{Institute for Quantum Information Processing, University of Ulm, D-89081 Ulm, Germany}
\author{Peter Zoller}
\affiliation{Institute for Quantum Optics and Quantum Information of the Austrian
Academy of Sciences\\
and Institute for Theoretical Physics, University of Innsbruck, A-6020
Innsbruck, Austria}

\title{Ion-assisted ground-state cooling of a trapped polar molecule}

\begin{abstract}
We propose and analyze a scheme for sympathetic cooling of the translational motion of polar molecules in an optical lattice, interacting one by one with laser-cooled ions in a radio-frequency trap. The energy gap between the excitation spectra of the particles in their respective trapping potentials is bridged by means of a parametric resonance, provided by the additional modulation of the RF field. We analyze two scenarios: simultaneous laser cooling and energy exchange between the ion and the molecule, and a scheme when these two processes take place separately. We calculate the lowest final energy of the molecule and the cooling rate depending on the amplitude of the parametric modulation. For small parametric modulation, the dynamics can be solved analytically within the rotating wave approximation.
\end{abstract}


\maketitle

\section{Introduction}

Ultracold molecules are the focus of increasing attention both in physics and in chemistry \cite{EPJD2004,Krems2005,Carr2009}. On one hand, recent experiments have succeeded in creating cold molecules via techniques like photoassociation of ultracold atoms \cite{Jones2006} or magnetically tunable Feshbach resonances \cite{Julienne2006}, and in cooling them to the ground vibrational state, e.g. via optical pumping \cite{Pillet2008,Denschlag2008,Ni2008,Danzl2010}. On the other hand, theoretical proposals have shown the relevance of such systems, especially heteronuclear molecules bearing nonzero electric dipole moment, for a range of applications from metrology \cite{Hinds2002,Demille2008} to quantum information processing \cite{Demille2002,Yelin2006} and the simulation of quantum many-body systems \cite{Wang2006,Micheli2006,Buchler2007}.

The preparation of trapped polar molecules in the ground state of both their internal rovibrational and center-of-mass motional states is therefore of crucial importance for further scientific and technological developments. Different approaches have been put forward, from the coupling to cavities, in the optical \cite{Morigi2007,Ritsch2008} as well as in the microwave regime \cite{Wallquist2008} to laser cooling \cite{Tannor2001,Robicheaux2009} and sympathetic cooling via collisions with neutral atoms \cite{Hutson2006,Doyle2007}. Cavity-based approaches, despite significant experimental requirements, hold the promise of reaching the ground state also for the molecular center-of-mass motion. On the other hand, buffer-gas cooling schemes based on neutral ``coolant'' atoms are so far limited in the reachable temperatures -- in the case of He \cite{Doyle2007}, simply by the buffer gas' own temperature, and in the case of Rb \cite{Hutson2006} by losses through inelastic collisions. Typically, different cooling schemes are applied to different temperature regimes of relevance for different types of experiments. The method we propose here is suitable for application in the ultracold regime in order to reach and maintain the molecular vibrational ground state via a sympathetic cooling procedure.

Our approach builds on recent advances in the combined manipulation of charged and neutral particles in AMO physics, where e.g. cold collisions between ions and neutral atoms have been observed \cite{Vuletic2009} and trapping of a single-ion within a Bose-Einstein condensate has been reported, including sympathetic cooling of the ion motion \cite{Zipkes2010}. The scenario we envision takes one step forward from these pioneering experiments and involves a one-dimensional array of molecules trapped in an optical lattice and interacting with a chain of ions in a nearby ion trap. We assume the molecules to be in the deep Mott-insulator phase, so that tunneling between neighboring lattice sites is strongly suppressed. In this regime, the problem can be reduced to considering pairwise interactions between one molecule and one ion at a time along the direction transverse to the trap axis.

In the following we will therefore focus on a single ion-molecule pair, and study in detail their interaction, with particular attention to their motional dynamics. Indeed, our idea for cooling the center-of-mass motion of a trapped molecule is based on a novel concept: the transfer of vibrational excitations to a neighboring, independently trapped ion that is subject to laser cooling. Repeated application of this procedure allows the molecule to reach the trap ground state over a time scale comparable to typical ion cooling rates. Moreover, the procedure is entirely insensitive to the molecule's internal state, and therefore is suitable for applications where the latter encodes information whose coherence needs to be preserved throughout motional cooling. The same applies, for instance, to atoms in an optical lattice used for quantum information processing, where quantum operations leave behind some residual heating of their external degrees of freedom. The ion-atom interaction shows the same distance dependence as the ion-molecule potential dealt with in this paper, and therefore our proposal is applicable also to the latter type of scenario, allowing to use one or more trapped ions to sympathetically cool the motion of qubit-carrying neutral atoms. This is relevant to ongoing experiments on ion-atom interactions that are being setup and carried out at several groups worldwide \cite{Makarov2005,Vuletic2009}.

The paper is structured as follows. In Sect.~\ref{Sec:Mod} we introduce the setup we have in mind and its description in terms of two harmonic oscillators coupled by the long-range ion-atom interaction, modeled according to a quantum-defect approach. In Sect.~\ref{Sec:Sep} we present the simplest scheme of alternating cooling and energy transfer stages by exploiting a parametric resonance obtained with trap frequency modulation. In Sect.~\ref{Sec:Sim} we discuss a different scheme in which the transfer and cooling processes take place simultaneously, so that a global master equation needs to be derived, which we do both within and beyond the rotating wave approximation in order to deal with arbitrary amplitudes of the parametric trap modulation. Finally, Sect.~\ref{Sec:Concl} contains a summary of our results and of future outlooks.

\section{The model\label{Sec:Mod}}

\subsection{Setup}

\begin{figure}
   \includegraphics[width=8cm,clip]{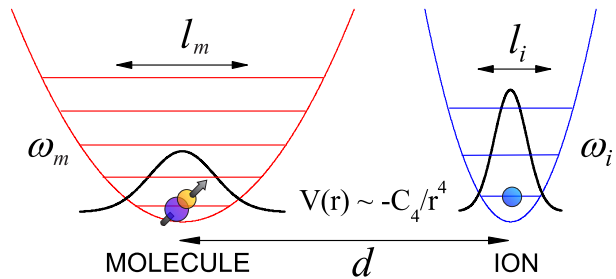}
     \caption{
     Specific considered setup consisting of a trapped polar molecule and an ion interacting via long-range polarization forces.
     \label{Fig:Setup}
   }
\end{figure}

As discussed above, we reduce the problem to considering a single spinless polar molecule in its electronic and rovibrational ground state, with zero projection of the total angular momentum on the internuclear axis ($X {}^1\Sigma, \nu = 0, J=0$). The molecule is confined in a far-off-resonance optical trap, created in a  laser standing wave (an optical lattice), and it interacts with a single ion confined in a radio-frequency (RF) trap (see Fig.~\ref{Fig:Setup}). Because of the different nature of the trapping potentials for the two particles, we can assume that the distance $d$ can be arbitrarily controlled by adjusting the external fields. A microscopic analysis of the ion-molecule system, presented in Appendix~\ref{App:Micro}, confirms that for typical parameters of the radio-frequency and far-off-resonance optical traps one can neglect the cross effects of the RF and optical fields on the molecule and ion, respectively. Moreover, the derivation shows that at large distances the ion attracts the polar molecule with a potential $V(r) \cong - C_4/r^4 $, ($r \rightarrow \infty$). Here, $C_4 = d_{m}^2 e^2/(6 B)$, where $d_{m}$ is the molecule permanent dipole moment, $e$ is the ion charge, and $B$ denotes the molecule rotational constant. An effective Hamiltonian for the system can be written as
\begin{align}
\label{H0}
H_0 = & \sum_{\nu = i,m} \left[ \frac{\mbf{p}_\nu^2}{2 m_\nu} +
\frac{1}{2} m_\nu \omega_\nu^2
\mbf{r}_\nu^2 \right]+ V(|\mbf{r}_i-\mbf{r}_m+\mbf{d}|),
\end{align}
where the label $i$ ($m$) refers to the ion (molecule), $\mbf{p}$ and $\mbf{r}$ are the momentum and position operators,
$\mbf{d}$ denotes the distance between the traps, and $m_m$, $\omega_m$ ($m_i$, $\omega_i$) are the molecule (ion) mass and trapping frequency, respectively. Without loosing generality we can assume that $\mbf{d}=d \mbf{e}_z$, and for simplicity we have assumed spherically symmetric trapping potentials.
%%%%%%%%%%%%%%%%%% Figure 1 %%%%%%%%%%%%%%%%%%%%%%
%%%%%%%%%%%%%%%%%%%%%%%%%%%%%%%%%%%%%%%%%%%%%%%%%%%%%

\subsection{Modeling as two coupled oscillators}

We assume that the distance $d$ between the two traps is sufficiently large, such that the effects of the ion-molecule interaction can be described perturbatively as distortions of the trapping potentials. In this regime the molecule and the ion remain well separated in the course of dynamics, and the ion-molecule interaction can be described by the asymptotic part of the potential: $V(r) = -C_4/r^4$. This allows also to neglect the effects of the tunneling through the potential barrier separating both particles, thus avoiding inelastic collisions. An expansion of $V(r)$ up to second order terms in the distance $r$ leads to
\begin{align}
\label{Happr}
H_0 \simeq & \sum_{\nu = i,a} \left[ \frac{\mbf{p}_\nu^2}{2 m_\nu} +
\frac{1}{2} m_\nu \bar{\omega}_\nu^2 (z_\nu-\bar{z}_\nu)^2 +
\frac{1}{2} m_\nu \tilde{\omega}_{\nu}^2 \bm{\rho}_\nu^2 \right] \nonumber \\
& + 20 \frac{C_4}{d^6} ({z_i}-\bar{z}_i)({z_m}-\bar{z}_m)-
4 \frac{C_4}{d^6} \bm{\rho}_m \bm{\rho}_i,
\end{align}
where $\bar{\omega}_{\nu}^2 = \omega_{\nu}^2 - 20 C_4/
(m_{\nu} d^6)$ and $\tilde{\omega}_{\nu}^2 = \omega_{\nu}^2 + 4 C_4/
(m_{\nu} d^6)$ are the trapping frequencies modified by the ion-molecule interaction,
$\bar{z}_\nu$ denotes the equilibrium position of the particles and $\bm{\rho} = (x,y,0)$. The  Hamiltonian (\ref{Happr})
describes a system of two coupled harmonic oscillators. Since the motion in the spatial directions $x$, $y$ and $z$
separates, in the following we focus on the dynamics along $z$, noting that a similar analysis can be carried out
for the $x$ and $y$ directions. Introducing the usual annihilation and creation operators:
\begin{align}
a & = \sqrt{\frac{m_i \bar{\omega}_i}{2\hbar}}\left[q_i + i \frac{(\mbf{p}_i)_z}{m_i \bar{\omega}_i}\right], \\
b & = \sqrt{\frac{m_m \bar{\omega}_m}{2\hbar}}\left[q_m + i \frac{(\mbf{p}_m)_z}{m_m \bar{\omega}_m}\right],
\end{align}
with $q_i = z_i - \bar{z}_i$, $q_m = z_m - \bar{z}_m$, $p_i = (\mbf{p}_i)_z$ and $p_m = (\mbf{p}_m)_z$, we rewrite the part of the Hamiltonian (\ref{Happr}) describing the motion in the $z$ direction in the following way:
\begin{equation}
\label{Hz}
H_{z} = \hbar \bar{\omega}_m a^\dagger a + \hbar \bar{\omega}_i b^\dagger b +
\hbar \omega_c (a + a^\dagger)(b + b^\dagger),
\end{equation}
where we omit the ground-state energy term. Here, $\omega_c$ denotes the coupling frequency:
$\hbar \omega_c = 10 C_4 l_m l_{i} /d^6$, and $l_{\nu}=\sqrt{\hbar/(m_\nu \bar{\omega}_\nu)}$, $\nu =i,m$ are the harmonic oscillator length for the ion and the molecule respectively. The value of $\omega_c$ is bounded by the condition that the quadratic terms in the expansion of the potential are positive, which yields the stable equilibrium position for both particles. This is fulfilled for $C_4 d^{-6} (1/m_i \omega_i^2+ 1/m_m \omega_m^2) \leq (\frac56)^6/20 \approx 0.017$, which brings some maximum value of the coupling frequency
\begin{equation}
\label{OmegaMax}
\omega_c^\mrm{max} = \frac12 \left( \frac{l_i}{\omega_i l_m}+ \frac{l_m}{\omega_m l_i}\right)^{-1}.
\end{equation}
For typical parameters for polar molecules in optical potentials and ions in RF traps, $\omega_c^\mrm{max}$ is of the order of $10$kHz. Considering an ultracold KRb molecule in the rovibrational singlet ground-state $X^1\Sigma$ ($d_m = 0.566$D and $B =1.1139$GHz \cite{Ni2008}) confined in an optical trap of frequency $\omega_m = 2 \pi \times 100 \textrm{kHz}$, and a ${}^{40}$Ca${}^+$ ion confined in an RF trap of frequency $\omega_i = 2 \pi \times 1\textrm{MHz}$, we obtain $\omega_c^\mrm{max} = 27.3$kHz corresponding to a trap separation $d =327$nm.

The Hamiltonian \eqref{Hz} can be diagonalized by applying the transformation described in Appendix~\ref{App:Transf}, which yields
\begin{equation}
\label{Hdiag}
H_z =\hbar \omega_1 \alpha^\dagger \alpha + \hbar \omega_2 \beta^\dagger \beta,
\end{equation}
where $\omega_1$ and $\omega_2$ are the frequencies of the two eigenmodes
\begin{align}
\label{omega12}
\omega_{1,2}^2 & = \frac{1}{2}\left[\bar{\omega}_i^2 + \bar{\omega}_m^2 \mp \sqrt{
(\bar{\omega}_i^2 - \bar{\omega}_m^2)^2 +
16 \omega_c^2 \bar{\omega}_m \bar{\omega}_i} \right],
\end{align}
with the upper (lower) sign referring to the frequency $\omega_1$ ($\omega_2$). For a weak coupling between the oscillators ($\omega_c \ll |\omega_i-\omega_m|$), the mode $\alpha^\dagger \alpha$ ($\beta^\dagger \beta$) is localized mainly on the molecule (ion). At distances when the coupled oscillator model remains valid, $\bar{\omega}_i \simeq \omega_i$ and $\bar{\omega}_m \simeq \omega_m$, and throughout the rest of the paper, we will neglect small corrections to the trapping frequencies due to the interaction potential, using the bare trapping frequencies $\omega_i$ and $\omega_m$.

\subsection{Quantum-defect treatment}
\label{Sec:QuantDef}

The validity of the coupled-oscillators model \eqref{Hz} can be verified by comparing it with a calculation based on a quantum-defect modeling of the short-range part of the interaction. For simplicity, we assume the same trapping frequencies for the molecule and for the ion: $\omega_i=\omega_m=\omega$,
which allows to decouple the center-of-mass (COM) and relative-motion degrees of freedom. We neglect the effects of the inelastic scattering that can occur in the collision of an ion and a ground-state ($X {}^1\Sigma, \nu = 0, J=0$) molecule, which are not important when the particles remain well separated in the course of the cooling process. The relative motion can be described by the single-channel Hamiltonian
\begin{align}
\label{Hrel}
H_\mrm{rel} = \frac{\mbf{p}^2}{2 \mu} +
\frac{1}{2} \mu \omega^2 (\mbf{r} - \mbf{d})^2 +V(r),
\end{align}
where we assume $V(r) =- C_4/r^4$ and we introduce some quantum-defect parameter $\varphi$, accounting for the effects of the short-range potential. A similar modeling turned out to be very successful in the case of ion-atom collisions \cite{AtomIon,IonMQDT}, where the interaction at large distances is governed by the same dispersion potential $V(r) = - C_4/r^4$. The quantum-defect parameter $\varphi$ can be interpreted as the short-range phase accumulated by the wave function due to the short-range
molecular potential. This interpretation follows from the behavior of the solution of the Schr\"odinger equation with the Hamiltonian \eqref{Hrel}. Performing the partial wave expansion $\Psi(r) = \sum_{l} R_l(r) Y_{lm}(\theta,\phi)$, we obtain the following short-range behavior of the radial wave function $R_l(r)$:
\begin{equation}
\label{Sol1}
R_l(r) \sim \sin\left(R^\ast/r + \varphi_l\right), \qquad r \ll \sqrt{R^{\ast}/k},R^{\ast}.
\end{equation}
Here $\hbar k$ is the relative momentum, $R^\ast \equiv \sqrt{2 C_4 \mu /\hbar^2}$ is some characteristic length scale for the molecule-ion interaction, and $\varphi_l$ is interpreted as a short-range phase. The asymptotic solution \eqref{Sol1} fulfills the radial Schr\"odinger equation neglecting the energy $\hbar^2 k^2/(2 \mu)$ and the centrifugal barrier $\hbar^2 l(l+1)/(2 \mu r^2)$ terms, which requires $r \ll \sqrt{R^{\ast}/k},R^{\ast}$. In this regime the phase $\varphi_l$ determined by the short-range interaction is independent of energy and angular momentum: $\varphi_l(E) \cong \varphi$. This allows to enclose the effects of the short-range forces into a single parameter $\varphi$, which together with Eq.~\eqref{Sol1} implies a boundary condition on the wave function at $r \rightarrow 0$. The fact that \eqref{Sol1} fulfills the Schr\"odinger equation for $k=0$ and $l =0$ allows to express the $s$-wave scattering length $a$ in terms of $\phi$: $a=-R^{\ast} \cot \phi$ \cite{AtomIon}. The parametrization in terms of $\varphi$ is valid as long as the distance $R_0$, at which $V(r)$ starts to deviate from the asymptotic $r^{-4}$ law (e.g due to the exchange interaction), is much smaller than $R^{\ast}$, which should be well fulfilled in the case of ion-atom and ion-molecule interactions.

\begin{figure}
   \includegraphics[width=7.8cm,clip]{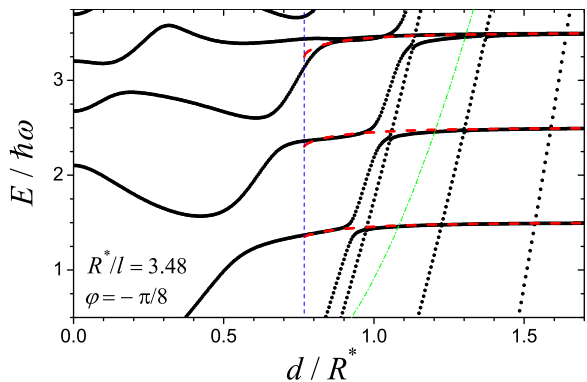}
	 \caption{
Energy spectrum of the relative motion for a molecule and an ion confined in spherically symmetric traps versus the distance $d$ between traps, calculated for some example parameters: $\varphi = -\pi/8$, and  $R^{\ast} = 3.48 l$, with $l=\sqrt{\hbar/\mu \omega}$. The exact numerical results (black lines) are compared with the prediction of the coupled-oscillator Hamiltonian \eqref{H2} (thick red dashed lines). The vertical blue dashed line shows the minimal distance at which the Hamiltonian can be expanded quadratically around the equilibrium points, whereas the green dash-dotted line shows the distance at which the barrier between particles becomes equal to the relative energy of the particles.
\label{Fig:spectrum3D} }
\end{figure}

In the framework of the quantum-defect model we can diagonalize the Hamiltonian \eqref{Hrel} with the boundary condition \eqref{Sol1} for small $r$, obtaining a set of eigenenergies $E_n(d)$. By comparing with the predictions of the coupled-oscillator model we can estimate the range of distances where the latter is still valid. Fig.~\ref{Fig:spectrum3D} shows the energy spectrum as a function of the trap separation $d$ obtained from the numerical diagonalization of \eqref{Hrel} for some example value of $\varphi$, together with a few lowest energy levels calculated from the coupled-oscillator approximation. The figure depicts also the distance at which the barrier between the particles becomes equal to the relative kinetic energy, and the minimal distance $d_\mrm{min}$ corresponding to $\omega_c^\mrm{max}$. We observe quite a good agreement between the oscillator model and the quantum-defect calculation; only close to $d_\mrm{min}$ does the oscillator model start to deviate from the accurate results. At such distances neglecting terms higher than quadratic in the expansion of the potential ceases to be valid. Apart from this, the two approaches differ in the vicinity of avoided crossings, representing trap-induced shape resonances \cite{AtomIon,Stock} between molecular and trap states, which obviously cannot be incorporated into the coupled-oscillator model.

%%%%%%%%%%%%%%%%%% Figure 2 %%%%%%%%%%%%%%%%%%%%%%
%%%%%%%%%%%%%%%%%%%%%%%%%%%%%%%%%%%%%%%%%%%%%%%%%%%%%

\subsection{Parametric resonance}

In general, the trapping frequencies for molecules trapped in the wells of an optical lattice are much smaller than the trapping frequency of an ion confined in an RF trap: $\omega_i \gg \omega_m$.
In this case the exchange of energy between the two particles is only possible in the presence of a time-dependent external potential, which can provide the difference in energies between the phonons.
In our scheme we propose to utilize the periodic modulation of the trapping frequency: $\omega^2(t) = \omega^2 (1 + 2 A \cos(\omega_\mrm{f} t))$, where $A$ is the amplitude of the modulation
and $\omega_\mrm{f}$ denotes its frequency. This can be achieved by applying an additional potential, which is
quadratic in space, and changes periodically in time. Periodic modulation of the trapping frequency
gives rise to the phenomenon of parametric resonance, allowing for exchange of energies between the particles.
In the subsequent analysis we consider the case when
the modulation is applied to the ion trap, however, the exchange of phonons can be also stimulated
by modulating the trapping frequency for the molecule, or the distance between the traps.

In the case of an ion stored in an RF trap, a periodic modulation of the trapping frequency can be realized by adding to the electric field a component varying with the frequency $\omega_\mrm{f}$. The stability of a single ion in such a potential is analyzed in Appendix~\ref{App:Stability}. In the presence of modulation of the trapping frequency for the ion $\omega^2_i(t) = \omega_i^2 (1 + 2 A \cos(\omega_\mrm{f} t))$, the total Hamiltonian $H$ takes the form
\begin{align}
\label{Hext}
H & = H_z + H_m, \\
\label{Hpar}
H_m & = \frac{A}{2} \hbar \omega_i \cos(\omega_\mrm{f} t) \left[c_\alpha(\alpha+\alpha^\dagger) +
c_\beta(\beta+\beta^\dagger) \right]^2.
\end{align}
Here, $c_\alpha$ and $c_\beta$ are the coefficients given in Appendix \ref{App:Transf}, resulting from the transformation diagonalizing the initial Hamiltonian. Now we
transform to the interaction picture with respect to $H_z$, and
sufficiently close to the resonance ($|\omega_2-\omega_\mrm{f}-\omega_1| \ll \omega_1,\omega_2$) we apply
the rotating-wave approximation (RWA), leaving in the transformed Hamiltonian $\tilde{H}_m= \exp(i H_z t/\hbar) H_m  \exp(- i H_z t/\hbar)$
only slowly varying terms:
\begin{equation}
\tilde{H}_m = A \hbar \omega_i c_\alpha c_\beta
[ \alpha \beta^\dagger \exp(i (\omega_\mrm{f}+\omega_1-\omega_2)t) + \mrm{h.c.}].
\end{equation}
Transforming back from the interaction
picture and moving to a rotating frame with respect to the second oscillator $\tilde{H} = \exp(- i H_2 t/\hbar) H
\exp( i H_2 t/\hbar)$, with $H_2 = \hbar \omega_\mrm{f} \alpha^\dagger \alpha$, we obtain
\begin{equation}
\label{HRWA}
\tilde{H}_\mrm{RWA} = \hbar (\omega_1 + \omega_\mrm{f}) \alpha^\dagger \alpha + \hbar \omega_2
\beta^\dagger \beta +\hbar \Omega_c (\alpha^\dagger \beta + \alpha \beta^\dagger).
\end{equation}
This Hamiltonian describes a system of two coupled, stationary oscillators with the coupling frequency
$\Omega_c = A \hbar \omega_i c_\alpha c_\beta$, which in the weak-coupling regime ($\omega_c \ll \omega_1, \omega_2,
|\omega_2-\omega_1|$) can be approximated by
\begin{equation}
\label{Omc}
\Omega_c \approx A \omega_c \omega_i^2 |\omega_i^2-\omega_m^2|^{-1}.
\end{equation}
Introducing the operators $T_1 = \alpha \beta^\dagger + \alpha^\dagger \beta$,
$T_2 = i(\alpha \beta^\dagger - \alpha^\dagger \beta)$ and $T_3 = \alpha^\dagger \alpha-\beta^\dagger \beta$, fulfilling the same commutation relations as the Pauli spin operators
\begin{equation}
\left[T_i,T_j \right] = 2 i \epsilon_{ijk} T_k,
\end{equation}
one immediately shows that (\ref{HRWA})
is equivalent to the Hamiltonian of a two-level system driven by an external
field. In the subspaces with
a constant total number of phonons: $N = \alpha^\dagger \alpha + \beta^\dagger \beta = \mrm{const}$, the Hamiltonian
(\ref{HRWA}) can be written as
\begin{equation}
\label{H2}
\tilde{H}_\mrm{RWA} = \frac{\hbar \Delta}{2} T_3 + \hbar \Omega_c T_1,
\end{equation}
where $\Delta = \omega_2 - \omega_1 - \omega_\mrm{f}$ and we omit $c$-number terms affecting only the evolution
of the total phase. Exactly at resonance ($\Delta = 0$), the expectation value of the difference of phonon numbers $\langle T_z \rangle $ evolves according to
\begin{equation}
\langle T_z \rangle(t) = \langle T_y \rangle(0) \sin (\theta(t)) +
\langle T_z \rangle(0) \cos (\theta(t)), \quad (\Delta=0)
\end{equation}
where $\theta(t) = 2 \int_0^t d\tau \Omega_c(\tau)$. Hence, application of the
$\pi$-pulse $2 \int_0^\infty d\tau \Omega_c(\tau)=\pi$ allows for exchange of the number of phonons between the molecule
and the ion, and, for instance, for realization of the SWAP gate based on qubits represented by motional states.
For a modulation frequency $\omega_\mrm{f}$ that is off-resonance,
and in the situation when initially all the phonon population is in one of the oscillators:
$\langle T_z \rangle(0) = N$, $\langle T_x \rangle(0) = \langle T_y \rangle(0) =0$,
the time-dependence of $\langle T_z \rangle$ is governed by
\begin{equation}
\label{Tz}
\langle T_z \rangle(t) = N \left[1 - \frac{8 \Omega_c^2}{4 \Omega_c^2+\Delta^2} \sin^2 \left(\frac{t\sqrt{4 \Omega_c^2+\Delta^2}}{2}\right)\right],
\end{equation}
where we assume that $\Omega_c(t)= \mrm{const}$. In this case the exchange of populations between the two oscillators is not complete, and one can observe that the decrease in the fidelity exchange has a typical Lorentzian shape with resonance width given by $2 \Omega_c$.

\begin{figure}
   \includegraphics[width=8.6cm,clip]{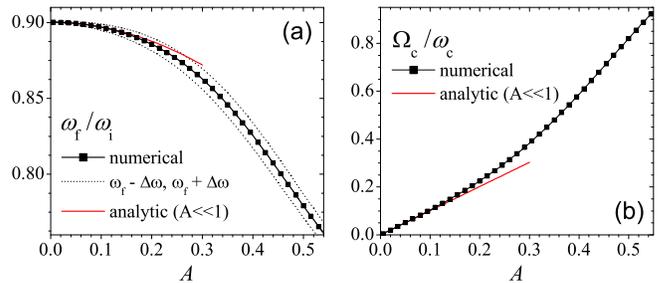}
	 \caption{
	 Panel (a):  modulation frequency $\omega_\mrm{f}$ that leads to the resonant coupling between the molecule and the ion (solid line with dots), and width of the resonance (dashed lines), calculated numerically from the continued fraction solution, compared with the analytical results (\ref{Res}) (solid line). Panel (b): effective coupling frequency $\Omega_c$ calculated numerically, compared with the analytical result of Eq.~(\ref{Omc}) (solid line). For both panels the calculations are performed with $\omega_m = 0.1 \omega_i$ and $\omega_c =
	 0.01 \omega_i$.
	 \label{Fig:Omf}
   }
\end{figure}

So far we have discussed the results obtained in RWA, which are valid for $A \ll 1$.
In general, for larger $A$ we can solve the time-dependent problem starting from the differential
equations for the position operators in the Heisenberg picture derived from the Hamiltonian \eqref{Happr}
\begin{equation}
\label{vectMath}
\mbf{\ddot{\hat{X}}} + B \mbf{\hat{X}} + 2 \cos(\omega_f t) C \mbf{\hat{X}} = 0,
\end{equation}
with the vector operator $\mbf{\hat{X}}^{T}(t) = (\hat{x}_i(t)-\bar{x}_i,\hat{x}_m(t)-\bar{x}_m)$, and
\begin{equation}
B= \left(
\begin{array}{ccc}
\omega_i^2 & 2 \omega_c \omega_i \\
2 \omega_c \omega_m & \omega_m^2
\end{array} \right), \quad
C= \left(
\begin{array}{ccc}
A \omega_i^2 & 0 \\
0 & 0
\end{array} \right).
\end{equation}
The form of Eq.~(\ref{vectMath}) resembles a Mathieu differential equation, and analogously its
solution can be found using Floquet expansion
%\begin{equation}
%\label{ansatz}
$\mbf{\hat{X}} = e^{i \lambda t} \sum_{n= - \infty}^{\infty} \mbf{\hat{v}}_n e^{i n \omega_{\! f} \, t}$.
%\end{equation}
Inserting this ansatz into (\ref{vectMath}) we find the recurrence relation for the operators
$\mbf{\hat{v}}_n$: $C(\mbf{\hat{v}}_{n+1} + \mbf{\hat{v}}_{n-1}) = D_n \mbf{\hat{v}}_n$, where
$D_n = I (\lambda + n \omega_f)^2 -B $, and $I$ is the unitary matrix. The recurrence relation can be solved
in terms of continued fractions: $\mbf{\hat{v}}_n = H_n \mbf{\hat{v}}_{n-1}$,
$\mbf{\hat{v}}_{-n} = H_{-n} \mbf{\hat{v}}_{-n+1}$, where $H_{n} = [D_{n} - C H_{n+1}]^{-1} C$,
$H_{-n} = [D_{-n} - C H_{-n-1}]^{-1} C$, and $n>0$. The eigenfrequencies $\lambda$ have to be determined from
the equation $\left[C\left(H_{1}(\lambda)+H_{-1}(\lambda)\right)-D_0 \right] \mbf{\hat{v}}_{0}=0$.
For $A \ll 1$, the lowest order approximation is obtained by assuming $H_{\pm 2}=0$,
which leads to the formula (\ref{Omc}) for $\Omega_c$ and gives the following resonance condition:
\begin{equation}
\label{Res}
\omega_2 - \omega_1 -\omega_f - \frac{A^2 \omega_i^3}{(\omega_i +\omega_m +
\omega_f)^2 -\omega_f^2} = 0.
\end{equation}
In comparison to the results obtained in
RWA, for nonzero $A$, the resonance is shifted by the presence of the modulated trapping
potential for the ion. This effect is illustrated on Fig.~\ref{Fig:Omf}.(a), showing
the position of the resonance as a function of $A$. It compares
the approximate result of Eq.~(\ref{Res}) with the full solution of (\ref{vectMath}) expressed in terms of continued fractions. In addition, Fig.~\ref{Fig:Omf}.(a)
presents the width of the resonance $\Delta \omega$, calculated by assuming a Lorentzian shape for the resonance peak \cite{footnote1}. We observe that the resonance is broader for larger values of the amplitude $A$, however for $A \gtrsim 0.3$
its width does not further increase.
The coupling frequency
$\Omega_c$ as a function of $A$ is presented in Fig.~\ref{Fig:Omf}.(b). This shows the analytical result (\ref{Omc})
and the exact result calculated from the continued fractions solution. We observe that for small $A$, $\Omega_c$
depends linearly on $A$, and it is well described by the analytical formula.
%%%%%%%%%%%%%%%%%% Figure 3 %%%%%%%%%%%%%%%%%%%%%%
%%%%%%%%%%%%%%%%%%%%%%%%%%%%%%%%%%%%%%%%%%%%%%%%%%%%%

\section{molecule cooling via parametric resonance}
\label{Sec:Sep}

\begin{figure}
   \includegraphics[width=8cm,clip]{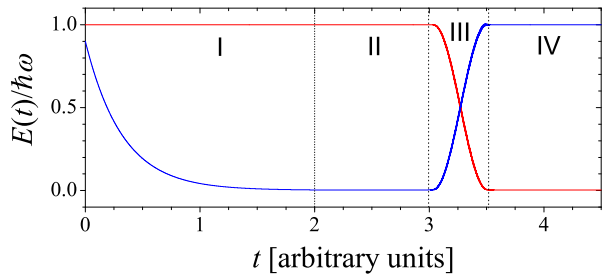}
     \caption{
     Typical dynamics of molecule and ion energies for the cooling scheme where the ion is cooled separately from the molecule. The roman number corresponds to the different cooling phases described in the text.
     \label{Fig:EnDep}
   }
\end{figure}

To utilize the molecule-ion interaction to cool the molecules, one can either consider collective laser-cooling in the coupled molecule-ion system, or apply laser cooling to the ion separately, and then with the help of modulation of the trapping frequency, exchange the energies of molecule and ion. In this section we focus on the latter scenario. We distinguish the following steps: (I) the ion is laser cooled;  (II) the particles are transferred to the distance allowing for efficient exchange of phonons between them; (III) the modulation of the trapping frequency is switched on and the energies of molecule and ion are exchanged by applying a $\pi$-pulse; (IV) the molecule and the ion are separated. %The lower panel shows the time-dependence of selected parameters (Rabi frequency $\Omega$ of the cooling laser, coupling frequency $\omega_c$, and amplitude $A$ of the parametric modulation) that control behavior of the system during the whole cooling process.
Fig.~\ref{Fig:EnDep} presents some typical dynamics of the cooling process. It shows the molecule and ion energies at different cooling steps, while consecutive cooling steps are separated by vertical lines.

%%%%%%%%%%%%%%%%%% Figure 4 %%%%%%%%%%%%%%%%%%%%%%
%%%%%%%%%%%%%%%%%%%%%%%%%%%%%%%%%%%%%%%%%%%%%%%%%%%%%

We first focus on the problem of molecule and/or ion transfer to the distances where the sympathetic cooling can be efficiently performed. In order to avoid additional heating due to the trap motion, we assume that the molecule and the ion traps are moved adiabatically, which requires (1) $\dot{\bar{z}}_k(t)/l_k \ll \omega_k(t)$ for $k=a,i$, (2) $|\dot{\omega}_j(t)| \ll \omega_1(t) \omega_c(t), \omega_2(t) \omega_c(t)$ for $j=1,2$. The former condition is necessary for the suppression of excitations during the movement of the traps, while the latter excludes the possibility of transitions due to the time dependence of $\omega_1(t)$ and $\omega_2(t)$, which change with the distance $d(t)$. In general the suppression of excitations due to the motion of the ion's trap can be achieved by fulfilling the weaker conditions \cite{Jaksch1999}
\begin{align}
\label{Transf1}
\left|\int_{t_0}^{t_1} \!\!\!\mrm{d}\tau \, \left[c_\alpha(\tau) f_i(\tau)+d_\alpha(\tau) f_m(\tau)\right] e^{i \phi_1(\tau)} \right| \ll 1, \\
\label{Transf2}
\left|\int_{t_0}^{t_1} \!\!\!\mrm{d}\tau \, \left[c_\beta(\tau) f_i(\tau)+d_\beta(\tau) f_m(\tau)\right] e^{i \phi_2(\tau)} \right| \ll 1,
\end{align}
with $c_\alpha$, $c_\beta$, $d_\alpha$, $d_\beta$ given in Appendix \ref{App:Transf}, and
\begin{align}
f_i(t) & = \omega_i \bar{z}_i(t)/l_i + 2 \omega_c(t) \bar{z}_m(t)/l_m, \\
f_m(t) & = \omega_m \bar{z}_m(t)/l_m + 2 \omega_c(t) \bar{z}_i(t)/l_i, \\
\phi_k(t) & = \int_{0}^{t} \!\!\!\mrm{d}\tau \omega_k(\tau), \quad  k=1,2,
\end{align}
where the limits $t_0$ and $t_1$ of the integral refer to the beginning and the end of the transfer process. Similarly, a weaker condition can be derived for the suppression of the probability of excitation due to the time dependence of $\omega_1(t)$ and $\omega_2(t)$:
\begin{align}
\label{Exc1}
&\left| \int_{t_0}^{t_1} \!\!\!\mrm{d}\tau \, \frac{\dot{\omega}_k(\tau)}{2 \omega_k(\tau)} e^{i 2 \phi_k(\tau)} \right| \ll 1, \quad k=1,2, \\
&\left| \int_{t_0}^{t_1} \!\!\!\mrm{d}\tau \, \frac{\dot{\omega}_1(\tau)+\dot{\omega}_2(\tau)}{4 \omega_c(\tau)} e^{i (\phi_1(\tau)+\phi_2(\tau))} \right| \ll 1,\\
\label{Exc3}
&\left| \int_{t_0}^{t_1} \!\!\!\mrm{d}\tau \, \frac{\dot{\omega}_1(\tau)-\dot{\omega}_2(\tau)}{4 \omega_c(\tau)} e^{i (\phi_1(\tau)-\phi_2(\tau))} \right| \ll 1,
\end{align}
In the phase (iii) when the energies of the molecule and of the ion are exchanged the pulse shape $A(t)$ should be optimized in order to obtain the best efficiency of the transfer, and to avoid excitations due to switching on and off the time-dependent modulation potential.

\begin{figure}
   \includegraphics[width=8.6cm,clip]{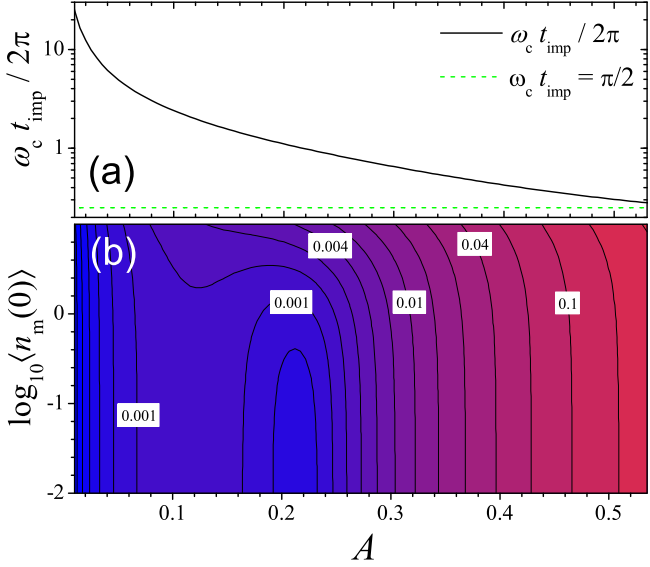}
	 \caption{
%	 Cooling of the molecule by ion, in the scheme when the exchange and cooling processes are separated.
	 Panel (a): Length of the rectangular $\pi$-pulse (solid line), and lower bound for the pulse length (dashed line) due to the physical coupling $\omega_c$.
	 Panel (b): Final mean phonon number of the molecule $\langle n_m \rangle$
	 as a function of the amplitude $A$ and of the
	 initial mean phonon number of the molecule $\langle n_m(0) \rangle$.
	 Calculations are performed with $\omega_m = 0.1 \omega_i$, $\omega_c =0.01 \omega_i$, and
	 $\langle n_{i}(0) \rangle = 0$.
	 \label{Fig:CoolSep}
   }
\end{figure}

We performed numerical simulations of the cooling process starting from different energies of the molecule before the cooling process, and assuming that the ion is laser-cooled to the ground state: $\langle n_{i}(0) \rangle = 0$. For simplicity we have assumed that the transfer process is ideal, i.e. the conditions \eqref{Transf1}-\eqref{Transf2} and \eqref{Exc1}-\eqref{Exc3} are fulfilled. In this way, after the transfer phase, the state of the particles is described by a density matrix $\rho = \rho_\alpha \otimes \rho_{\beta}$, where $\rho_\alpha=\sum_n p_n |n\rangle\langle n|$ and $\rho_\beta=|0\rangle\langle 0|$ correspond, respectively, to the state of the first and the second harmonic oscillator of the Hamiltonian~\eqref{Hdiag}. Fig.~\ref{Fig:CoolSep} shows the final mean phonon number of the molecule
for $\omega_m = 0.1 \omega_i$, $\omega_c = 0.01 \omega_i$. Fig.~\ref{Fig:CoolSep}.(a) shows the length $t_{i}$ of the exchange pulse as a function of the amplitude $A$, while the dashed line
presents the lower bound for the length due to the physical coupling $\omega_c$ between the oscillators: $t_\mrm{min} = \pi/(2 \omega_c)$. For simplicity, in the calculations we have assumed a rectangular shape of the pulse for the amplitude of the parametric modulation $A(t)$. Hence the pulse length is given by $t_\mrm{imp} = \pi/(2 \Omega_c)$. The sudden turning on and off of the parametric modulation leads to an additional heating that increases with the amplitude $A$. This is shown in Fig.~\ref{Fig:CoolSep}.(b), presenting the final mean phonon number of the molecule $\langle n_m \rangle$ for different values of the initial mean phonon number $\langle n_m(0) \rangle$.
We observe that, even without using an optimized pulse for $A(t)$, the heating introduced by imperfections of the transfer process remains quite small for $A \lesssim 0.3$, and is practically independent of  $\langle n_m(0) \rangle$. We notice that the increase of the molecule energy with $A$ is not monotonic and that there is a minimum in the molecule energy around $A \approx 0.2$. On the contrary, the molecule and ion energies for detunings far from the parametric resonance, when the exchange of energy is suppressed, increase monotonically with $A$.

%%%%%%%%%%%%%%%%%% Figure 5 %%%%%%%%%%%%%%%%%%%%%%
%%%%%%%%%%%%%%%%%%%%%%%%%%%%%%%%%%%%%%%%%%%%%%%%%%%%%

\section{Simultaneous laser cooling and energy exchange}
\label{Sec:Sim}

We turn now to the description of laser cooling in a coupled system consisting of a trapped molecule and an ion, where laser cooling acts at the same time as the parametric modulation. This process consists of three separate phases, (I) transfer of particles to distances allowing for efficient exchange of phonons between the molecule and the ion; (II) laser cooling and simultaneous exchange of energy through parametric resonance; (III) separation of the particles.
%In the lower panel we show again the time dependence of some selected parameters that control the behavior of the system during the individual cooling phases.

\subsection{Master Equation}

We assume that the cooling laser affects only the ion, and that the coupling between molecule and ion
is sufficiently weak: $\omega_c \ll \Omega, \gamma$, where $\Omega$ is the Rabi frequency due to the interaction of the ion with the laser field, and $\gamma$ is the spontaneous emission rate. In this way, the absorption and spontaneous emission processes of the ion are not affected by the coupling with the molecule. Similarly to the considerations presented in the previous sections, we restrict ourselves to the case of one-dimensional
motion along the direction determined by the centers of the traps.
Using the standard theory of quantum stochastic processes \cite{Gardiner,Carmichael},
we derive a master equation (ME) for the reduced density operator $\rho$ obtained
by tracing over the empty modes of the radiation field. In the frame rotating at the laser frequency $\omega_L$, the
ME reads \cite{Cirac}
\begin{equation}
\label{ME}
\begin{split}
\dot{\rho} = & \frac{1}{i \hbar} \left[ H + H_{int} + H_{las},\rho \right] \\
& + \frac{\gamma}{2} \bigg[ \sigma_{-} \int_{-1}^{1} du \, {\cal W}(u) e^{i k z_i u} \rho
e^{- i k z_i u} \sigma_{+} \\
& \qquad \quad  - \sigma_{+} \sigma_{-} \rho - \rho \sigma_{+} \sigma_{-} \bigg].
\end{split}
\end{equation}
Here, $H$ denotes the Hamiltonian (\ref{Hext}) for the external degrees of freedom,
$H_{int}$ is the Hamiltonian for the internal degrees of freedom
\begin{equation}
H_{int} = - \frac{\hbar \Delta}{2} \sigma_z,
\end{equation}
where $\Delta = \omega_L - \omega_0$ denotes detuning, $\omega_0$ is the frequency of the optical transition,
and $\sigma_z, \sigma_{\pm}$ are the standard operators describing transitions in the two-level molecule.
The Hamiltonian $H_{las}$ describes the interaction of the ion with the laser field
\begin{equation}
H_{las} = \frac{\hbar \Omega}{2} \left( e^{i k z_i} \sigma_{+} + h.c. \right),
\end{equation}
with $k$ denoting the wave vector of the laser field. Finally
${\cal W}(u)$ is the angular pattern of the spontaneous
emission, which for dipolar transitions is ${\cal W} = \frac{3}{4} (1+ u^2)$.

We focus now on cooling in the Lamb-Dicke regime, where the laser wavelength is much larger than the ion's
trap size. The presence
of the small Lamb-Dicke parameter $\eta = k \sqrt{\hbar/(2 m_i \omega_i)}$,
allows to expand the ME in powers of $\eta$.
To lowest order in $\eta$ the evolution of the internal and external degrees of freedom is decoupled, and
the internal state can be adiabatically eliminated (see Appendix \ref{App:AdiabElim}). The
adiabatic elimination is valid provided that the internal dynamics is faster than the lowest-order
coupling between the internal and external degrees of freedom, which requires
$\eta \Omega \ll \gamma,\Omega$. As a result of the adiabatic elimination, we obtain the following ME
for the reduced density
operator $\rho_e$ obtained by tracing over the internal degrees of freedom
\begin{equation}
\label{MEae}
\begin{split}
\dot{\rho}_e = & \frac{1}{i \hbar} \left[ H
- \frac{\hbar \Omega}{2} \langle \sigma_y \rangle_{s} k z_i
- \frac{\hbar \Omega}{4} \langle \sigma_x \rangle_{s} k^2 z_i^2
,\rho_e \right] \\
& + \frac{\gamma}{5} k^2 \langle \sigma_{+} \sigma_{-} \rangle_{s}
\left(2 z_i \rho_e z_i - z_i^2 \rho_e - \rho_e z_i^2 \right) \\
& + \frac{k^2 \Omega^2}{4} \int_0^{\infty} \! \! \! d\tau \,  \bigg[
\left( z_i(\tau) \rho_e z_i - z_i z_i(\tau) \rho_e \right) \\
& \qquad \qquad \qquad \quad \times \Big(\langle \sigma_y(\tau) \sigma_y(0) \rangle_{s}
- \langle \sigma_y \rangle_{s}^2 \Big) + h.c. \bigg],
\end{split}
\end{equation}
where we include terms up to the second order in $\eta$. Here,
\begin{gather}
z_i(\tau)  = e^{\tau {\cal L}_{0E}} z_i = e^{- i t H /\hbar} z_i
e^{i t H/\hbar}, \\
\langle \sigma_y(\tau) \sigma_y(0) \rangle_{s} =
\textrm{Tr}_I \{ \sigma_y e^{\tau {\cal L}_{0I}} \sigma_y \rho_{ss} \}.
\end{gather}
and $\langle . \rangle_{s} \equiv \textrm{Tr}_I \{ . \rho_{s} \}$ denotes an average taken in the stationary state of
the density matrix $\rho_{s}$ of the internal degrees of freedom. It is defined by ${\cal L}_{0I} \rho_{s} =0$,
where the Liouvillians ${\cal L}_{0I}$ and ${\cal L}_{0E}$ are given by
\begin{align}
{\cal L}_{0I} \rho =
&  \frac{1}{i \hbar}
\left[  \frac{\hbar \Omega}{2} (\sigma_{-}+\sigma_{+}) - \frac{\hbar \Delta}{2} \sigma_z,\rho \right] \nonumber \\
& + \frac{\gamma}{2} \left[ 2 \sigma_{-} \rho \sigma_{+} - \sigma_{+} \sigma_{-} \rho -
\rho \sigma_{+} \sigma_{-} \right], \\
{\cal L}_{0E} \rho = & \frac{1}{i \hbar} \left[ H, \rho \right].
\end{align}
The two-time correlation functions $\langle \sigma_y(\tau) \sigma_y(0)
\rangle_{ss}$ can be calculated with the help of the quantum regression theorem (see Appendix \ref{App:TTC} for details).
The first term of the ME (\ref{MEae}) describes the evolution due to
the trapping potentials and the coupling between the molecule and the ion. We observe that the trapping potential for the ion is
modified by the interaction with the laser field. The second term of (\ref{MEae}) represents damping due to
spontaneous emission, while the third term describes the excitation of the ion due the cooling laser.

\subsection{Results in RWA}

In this section we discuss laser cooling in the molecule-ion system on the level of RWA. Although a treatment based on RWA
is in principle limited to the regime $A \ll 1$, the analysis presented in this section provides a good qualitative
description of the cooling dynamics, and it can be carried out fully analytically.
A description of the laser
cooling for arbitrary $A$ requires going beyond RWA, and this is addressed in the next section.

In the framework of RWA we make the following assumptions: (i) $H$ is approximated by $\tilde{H}_{RWA}$ given by
\eqref{HRWA}; (ii) RWA is applied to the second and to the third term in the ME \eqref{MEae}, which results in omitting
the quickly rotating terms;
(iii) we neglect the second and the third term in the commutator in \eqref{MEae}, describing the effects of
the laser field on the ion's trap, which requires $\eta \Omega \ll \omega_i$. In this way the ME \eqref{MEae}
reduces to
\begin{equation}
\label{MERWA}
\begin{split}
\dot{\rho_e} = &  \frac{1}{i \hbar} \left[ \tilde{H}_{RWA},\rho_e \right] + \\
& + \bigg\{ S_{a}^{-} (\alpha \rho_e \beta^\dagger - \beta^\dagger \alpha \rho_e)
+ S_{a}^{+} (\alpha^\dagger \rho_e \beta - \beta \alpha^\dagger \rho_e) \\
& \quad \, + (S_{b}^{-} + D) (\beta \rho_e \beta^\dagger - \beta^\dagger \beta \rho_e) \\
& \quad \, + (S_{b}^{+} + D)(\beta^\dagger \rho_e \beta - \beta \beta^\dagger \rho_e) + h.c. \bigg\},
\end{split}
\end{equation}
where $D = \eta^2 \gamma \langle \sigma_{+} \sigma_{-} \rangle_{s}/5$,
\begin{align}
\label{S-}
S_\nu^{-} & =  \frac{\eta^2 \Omega^2}{4} \int_{0}^\infty \!\! d\tau \,
f_\nu^{\ast}(\tau) \left( \langle \sigma_y(\tau) \sigma_y(0) \rangle_{s} - \langle \sigma_y \rangle_{s}^2 \right), \\
\label{S+}
S_\nu^{+} & =  \frac{\eta^2 \Omega^2}{4} \int_{0}^\infty \!\! d\tau \,
f_\nu(\tau) \left( \langle \sigma_y(\tau) \sigma_y(0) \rangle_{s} - \langle \sigma_y \rangle_{s}^2 \right),
\end{align}
for $\nu=a,b$. Here $f_\nu(\tau)$ are the functions determined by the evolution of the operator $\beta$ in the
Heisenberg picture
\begin{equation}
\label{betatau}
\beta(\tau) = e^{i t \tilde{H}_{RWA} /\hbar} \beta  e^{-i t \tilde{H}_{RWA}/\hbar} =
f_m(\tau) \alpha + f_b(\tau) \beta.
\end{equation}
We observe that the ME \eqref{MERWA} contains, apart from the terms describing the laser cooling of ion, also mixed
terms involving $\alpha$ and $\beta$ operators corresponding to the action of the laser field on the molecule.
In the considered regime of weak coupling $\omega_c \ll \gamma,\Omega$, mixed terms can be neglected in
comparison to the terms describing the laser cooling of the ion only. This can be verified
by inserting the solutions for $f_m(\tau)$ and $f_b(\tau)$, which
exactly at parametric resonance ($\omega_1+ \omega_\mrm{f} = \omega_2$) takes the form
$f_m(\tau) = -i e^{- i \omega_2 t} \sin (\Omega_c t)$ and
$f_b(\tau) = e^{- i \omega_2 t} \cos (\Omega_c t)$. Inserting these solutions into \eqref{S-}-\eqref{S+} we obtain
\begin{align}
S_a^{-} & = {\tst \frac12} \left[S(\omega_2 + \Omega_c) - S(\omega_2 - \Omega_c)\right], \\
S_b^{-} & = {\tst \frac12} \left[S(\omega_2 + \Omega_c) + S(\omega_2 - \Omega_c)\right], \\
S_a^{+} & = {\tst \frac12} \left[S(-\omega_2 - \Omega_c) - S(-\omega_2 + \Omega_c)\right], \\
S_b^{+} & = {\tst \frac12} \left[S(-\omega_2 - \Omega_c) + S(-\omega_2 + \Omega_c)\right],
\end{align}
where
\begin{equation}
\label{Somega}
S(\omega) =  \frac{\eta^2 \Omega^2}{4} \int_{0}^\infty \!\! d\tau \,
e^{i \omega \tau} \left( \langle \sigma_y(\tau) \sigma_y(0) \rangle_{s} - \langle \sigma_y \rangle_{s}^2 \right),
\end{equation}
The function $S(\omega)$ is calculated in Appendix \ref{App:TTC}, and from the form of this solution we observe that
$S(\omega)$ is determined by the laser cooling parameters $\gamma$, $\Omega$, and $\Delta$. Therefore
in the regime $\Omega_c \ll \gamma,\Omega,\Delta$ we have $S(\omega_2 + \Omega_c)  \approx S(\omega_2 - \Omega_c)
\approx S(\omega_2)$ and $S(-\omega_2 + \Omega_c)  \approx S(-\omega_2 + \Omega_c)
\approx S(-\omega_2)$. Hence, in the lowest approximation the coefficients $S_b^{-}$ and $S_b^{+}$ becomes independent of
the coupling between molecule and ion, and identical with the
coefficients describing the laser cooling of a single ion.
On the contrary,
in this regime the coefficients $S_a^{-}$ and $S_a^{+}$ become very small, therefore the terms
describing the damping process
and involving $\alpha$ and $\alpha^\dagger$ operators can be neglected.

\begin{figure}
   \includegraphics[width=6.5cm,clip]{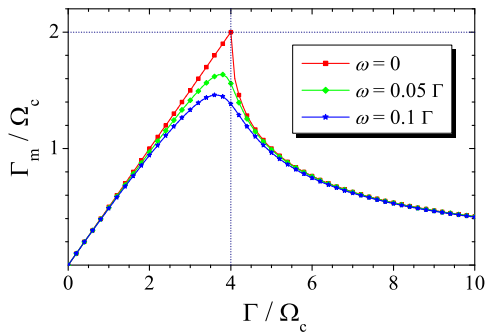}
	 \caption{
	 Cooling rate of the molecule $\Gamma_m$ as a function of the cooling rate of the ion $\Gamma$,
	 for different values of the parameter $\omega$ (see text for details). Cooling rates are scaled by the
	 effective coupling $\Omega_c$.
	 \label{Fig:Gamma}
   }
\end{figure}

\begin{figure}
   \includegraphics[width=8.6cm,clip]{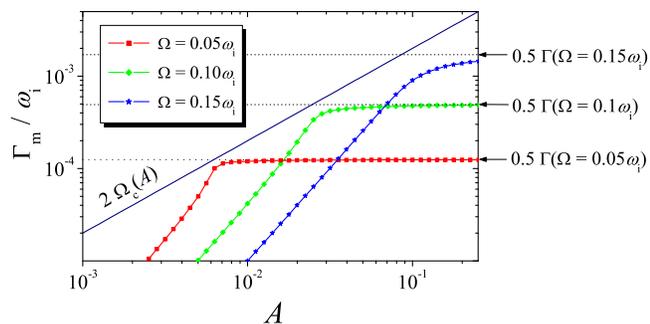}
	 \caption{
	 Cooling rate of the molecule $\Gamma_m$ versus the amplitude of the parametric modulation $A$, for different values of the
	 Rabi frequency $\Omega$. For comparison we include a line showing the values of $2 \Omega_c$ (solid line
	 without symbols), and the cooling rate of the ion $\Gamma$ (horizontal lines).
	 The results are obtained for $\omega_m = 0.2 \omega_i$ and $\omega_c =0.01 \omega_i$, $m_m= m_i$ and for
	 cooling parameters corresponding to the sideband cooling regime: $\gamma=0.1 \omega_i$, $\Delta= - \omega_i$ and
	 $\eta =0.1$.
	 \label{Fig:GammaA}
   }
\end{figure}

From the ME (\ref{MERWA}) we derive a set of differential equations describing the dynamics of the cooling process.
This straightforward calculations leads to
\begin{eqnarray}
\label{u1_ev}
\dot{u} & = & i \omega v + (a_{+} - a_{-}) u , \\
\dot{v} & = & i \omega u + (a_{+} - a_{-}) v - i 2 \tilde{\omega}_c (w-s) ,\\
\dot{s} & = & i \tilde{\omega_c} v + 2(a_{+} - a_{-}) s , \\
\label{w1_ev}
\dot{w} & = & - i \tilde{\omega}_c v,
\end{eqnarray}
where the variables $u$, $v$, $s$ and $w$ are defined by
$s = \langle \beta^\dagger \beta\rangle
- \langle \beta^\dagger \beta\rangle_{s}$, $w = \langle \alpha^\dagger \alpha\rangle
- \langle \beta^\dagger \beta\rangle_{s}$, $u = \langle \beta \alpha^\dagger\rangle +
\langle \beta^\dagger \alpha \rangle$, $v = \langle \beta^\dagger \alpha \rangle
- \langle \beta \alpha^\dagger\rangle$. Here $\langle \beta^\dagger \beta\rangle_{s} = a_{+}/(a_{-} - a_{+})$
denotes the final energy of the ion at the end of the cooling process,
$a_{\pm} = \textrm{Re}(S_b^{\pm} + D)$, $\omega_{\pm} = \textrm{Im}(S_b^{\pm} + D)$ and
$\omega = \omega_2 - \omega_1 - \omega_\mrm{f} + \omega_{+} + \omega_{-}$. We observe that the final
(equilibrium) energy of the molecule is the same as for the ion in the absence of the molecule \cite{Cirac}.
The set of differential equations
(\ref{u1_ev})-(\ref{w1_ev}) has one pair of complex eigenvalues and one pair of real eigenvalues. One can
easily verify that the larger of the two real eigenvalues provides the cooling rate of the molecule
\begin{equation}
\label{GA}
\begin{split}
\Gamma_m = & - \frac{\Gamma}{2} + \frac{1}{\sqrt{2}} \bigg[ \Big( \big( \omega^2 + 4 \Omega_c^2 - {\tst \frac14}
\Gamma^2\big)^2 + \omega^2 \Gamma^2 \Big)^{\!1/2} \\
& + {\tst \frac14} \Gamma^2 - 4 \Omega_c^2 - \omega^2 \bigg],
\end{split}
\end{equation}
where $\Gamma = 2 (a_{-} - a_{+})$ is the cooling rate of the ion in the absence of the molecule.
At the parametric resonance we have $\omega = \omega_{+} +
\omega_{-}$, and for typical parameters corresponding to the sideband cooling regime we get
$\omega_{+},\omega_{-} \ll \Gamma$. The cooling rate of the molecule $\Gamma_m$ as a function of the cooling rate of the ion
$\Gamma$, for different values of the parameter $\omega$ is shown in Fig.~\ref{Fig:Gamma}. We note that for
$\Gamma \lesssim 4 \Omega_c$ the molecule is cooled with a rate $\Gamma/2$. Further increasing of $\Gamma$ does not result
in faster cooling of the molecule, and in the regime $\Gamma \gtrsim 4 \Omega_c$ the cooling process becomes ineffective.
This behavior is also illustrated in Fig.~\ref{Fig:GammaA}, showing the cooling rate versus the amplitude $A$ of
the parametric modulation, calculated for different values of the Rabi frequency $\Omega$. The presented results
are obtained for the following values of the parameters:
$\omega_m = 0.2 \omega_i$, $\omega_c =0.01 \omega_i$, $m_m= m_i$, $\gamma=0.1 \omega_i$, $\Delta= - \omega_i$
and $\eta = 0.1$. For comparison we include a line showing the values of
$2 \Omega_c$. We observe that for too small values of $A$ (when $4 \Omega_c \lesssim \Gamma$)
the cooling is inefficient and the cooling rate of
the molecule is smaller both than $\Gamma$ and $\Omega_c$. For larger values of $A$, when
$4 \Omega_c \gtrsim \Gamma$,
the cooling rate $\Gamma_m$ becomes independent of $A$ and equal to $\Gamma/2$.
%%%%%%%%%%%%%%%%%% Figure 6 %%%%%%%%%%%%%%%%%%%%%%
%%%%%%%%%%%%%%%%%%%%%%%%%%%%%%%%%%%%%%%%%%%%%%%%%%%%%
%%%%%%%%%%%%%%%%%% Figure 7 %%%%%%%%%%%%%%%%%%%%%%
%%%%%%%%%%%%%%%%%%%%%%%%%%%%%%%%%%%%%%%%%%%%%%%%%%%%%

\subsection{Beyond RWA}

\begin{figure}
   \includegraphics[width=8.6cm,clip]{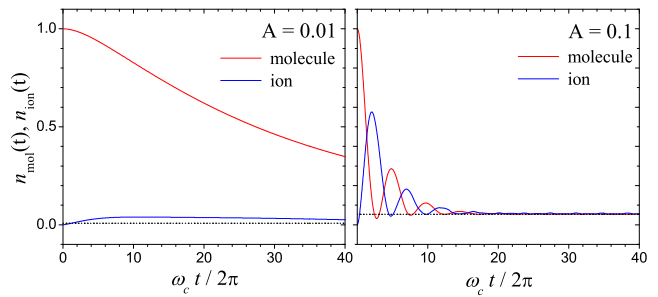}
	 \caption{
	 Dynamics of the
	 mean phonon number of the molecule $n_\mrm{mol}(t)$ and the ion $n_\mrm{ion}(t)$. The calculations are performed for the parameters of Fig.~\ref{Fig:GammaA} and $\Omega = 0.1 \omega_i$.
	 The dotted horizontal line shows the final energy for the molecule and the ion.
	 \label{Fig:Dyn}
   }
\end{figure}

In this section we consider cooling in the ion-molecule system taking into account effects beyond RWA.
We start from the ME (\ref{MEae}) with $H$ represented in terms of momentum and position operators
\begin{equation}
\label{H}
\begin{split}
H = & \sum_{\nu = i,a} \left[ \frac{p_\nu^2}{2 m_\nu} +
\frac{1}{2} m_\nu \omega_\nu^2
z_\nu^2 \right] + \frac{2 \hbar \omega_c}{l_i l_m} z_i z_m \\
& + A m_i \omega_i^2 z_i^2 \cos (\omega_\mrm{f} t),
\end{split}
\end{equation}
and we derive equations of motion for average values quadratic in the operators
$\hat{z}_i$, $\hat{p}_i$, $\hat{z}_m$, $\hat{p}_m$. In principle the term linear in $\hat{z}_i$ in the commutator in
(\ref{MEae}) couples the observables quadratic in positions or momenta to the linear ones. However,
similarly to the analysis
performed in RWA, we can neglect it assuming the effects of the laser field on the trapping potential for the ion are small for
$\eta \Omega \ll \omega_i$. In this way we obtain a set of linear equations, which we discuss in
Appendix~\ref{App:BeyRWA}. We solve the equations of motion numerically, determining the dynamics of the cooling
process and the final energies of the molecule and the ion.
Some sample dynamics of the cooling process is presented in Fig.~\ref{Fig:Dyn}, which depicts the mean
phonon number of the molecule and of the ion as a function of time.
The presented results are calculated for $\omega_m = 0.2 \omega_i$, $\omega_c =0.01 \omega_i$, $m_m= m_i$,
and for cooling parameters corresponding to the regime of sideband cooling:
$\gamma=\Omega=0.1 \omega_i$, $\Delta= - \omega_i$ and $\eta = 0.1$.
Fig.~\ref{Fig:Dyn}.(a) presents the dynamics for $A=0.01$,
for which $\Gamma \gtrsim 4 \Omega_c$ (cf. Fig.~\ref{Fig:GammaA}). In this case we observe a very weak exchange of energies between the particles, which is due to the too fast cooling of the ion. This behavior is somehow similar to the Zeno effect that can be observed in quantum systems, corresponding to the suppression of the dynamics in the presence of frequent measurements performed on the system. In our case the dynamics of energy exchange is suppressed due to the too strong damping of the ion's motion. On the other hand, for $A=0.1$ the parameters are well within the regime of efficient cooling ($\Gamma \lesssim 4 \Omega_c$), which can be deduced from Fig.~\ref{Fig:GammaA}. In this case the dynamics
exhibits the presence of oscillations,
which are damped in time due to the laser cooling process,
which is illustrated in Fig.~\ref{Fig:Dyn}.(b).
%%%%%%%%%%%%%%%%%% Figure 8 %%%%%%%%%%%%%%%%%%%%%%
%%%%%%%%%%%%%%%%%%%%%%%%%%%%%%%%%%%%%%%%%%%%%%%%%%%%%

\begin{figure}
   \includegraphics[width=8.6cm,clip]{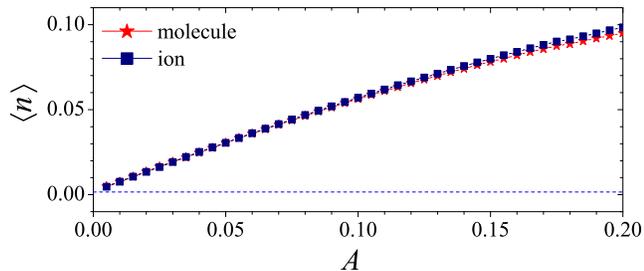}
	 \caption{
	 The final mean phonon number of the molecule and the ion as a function of the amplitude $A$.
	 The calculations are performed for the same parameters as in Fig.~\ref{Fig:Dyn}.
	 \label{Fig:Efin}
   }
\end{figure}

Fig.~\ref{Fig:Efin} shows the dependence of the final energy of the molecule and the ion
on the amplitude $A$, calculated for the same parameters as in Fig.~\ref{Fig:Dyn}.
We observe that the final energies of the molecule and the ion are equal,
and that they increase with the amplitude $A$. Hence, the
presence of the oscillating potential decreases the efficiency of laser cooling. To investigate whether this
effect is related with the coupling between the molecule and the ion, or if it originates purely from the parametric modulation
of the ion's trap, we performed simulations of the cooling, putting $\omega_c=0$, which describes the situation
of the molecule separated from the ion.
%%%%%%%%%%%%%%%%%% Figure 9 %%%%%%%%%%%%%%%%%%%%%%
%%%%%%%%%%%%%%%%%%%%%%%%%%%%%%%%%%%%%%%%%%%%%%%%%%%%%

\begin{figure}
   \includegraphics[width=7cm,clip]{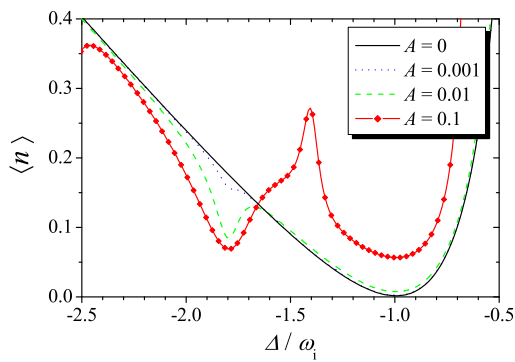}
   \caption{
   The final energy of the ion as a function of the detuning $\Delta$, and for different values of
   the amplitude $A$. The results are calculated for the parameters of Fig.~\ref{Fig:GammaA}.
	 \label{Fig:EionD}
	 }
\end{figure}

Fig.~\ref{Fig:EionD} shows the dependence of the final energy of the ion on the detuning
$\Delta$, calculated for different values of the amplitude $A$.
We observe that the modulation of the trapping frequency leads to an increase in the final energy,
similarly to what is observed in the coupled ion-molecule system.
For the parameters used in the simulation, the cooling
is performed in the resolved-sideband regime, and for non-zero
values of $A$ we observe the appearance of a second sideband at the frequency $\omega_i + \omega_f$.
For higher values of $A$ ($A \sim 0.1$),
the cooling tuned to this second sideband is almost as efficient as the cooling at the sideband located
at $\omega_i$.
%%%%%%%%%%%%%%%%%% Figure 10 %%%%%%%%%%%%%%%%%%%%%%
%%%%%%%%%%%%%%%%%%%%%%%%%%%%%%%%%%%%%%%%%%%%%%%%%%%%

\begin{figure}
   \includegraphics[width=7cm,clip]{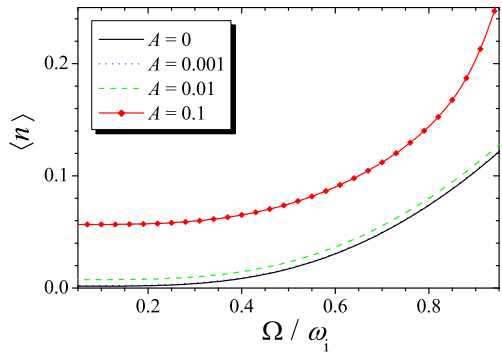}
   \caption{
   The final energy of the ion as a function of the Rabi frequency $\Omega$, and for different values of
   the amplitude $A$. The results are calculated for the parameters of Fig.~\ref{Fig:GammaA}.
	 \label{Fig:EionOm}
	 }
\end{figure}

\begin{figure}
   \includegraphics[width=7cm,clip]{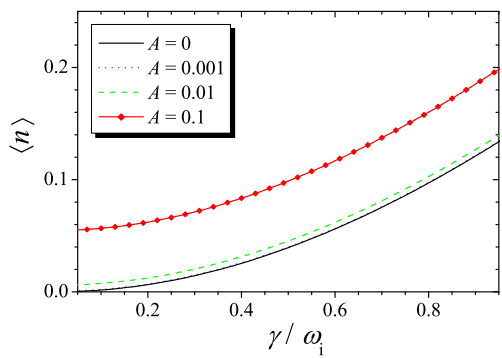}
   \caption{
   The final energy of ion as a function of the spontaneous emission rate $\gamma$, and for different values of
   the amplitude $A$. The results are calculated for the parameters of Fig.~\ref{Fig:GammaA}.
	 \label{Fig:EionG}
	 }
\end{figure}

Figs.~\ref{Fig:EionOm} and \ref{Fig:EionG} shows the final energy of the ion as a function
of the Rabi frequency and of the spontaneous emission rate, respectively,
evaluated for different values of the amplitude $A$. Again, we note that for any combination of the cooling parameters
the final energy of the ion increases with $A$. Finally,
we can conclude that the lowest energies are achieved when $\Omega,\gamma \ll \omega_i$,
and for the laser tuned to the first red-detuned
sideband $\Delta = - \omega_i$. At the same time, the amplitude $A$ should
be kept as small as possible, limited only by desired rate of the energy transfer between the molecule and the ion.
%%%%%%%%%%%%%%%%%% Figure 11 %%%%%%%%%%%%%%%%%%%%%%
%%%%%%%%%%%%%%%%%%%%%%%%%%%%%%%%%%%%%%%%%%%%%%%%%%%%
%%%%%%%%%%%%%%%%%% Figure 12 %%%%%%%%%%%%%%%%%%%%%%
%%%%%%%%%%%%%%%%%%%%%%%%%%%%%%%%%%%%%%%%%%%%%%%%%%%%

\section{Conclusions\label{Sec:Concl}}

In summary, we have analyzed sympathetic cooling of the center-of-mass motion of a polar molecule prepared in the internal rovibrational ground state and interacting with a laser-cooled ion. The ion and the molecule are confined in separate trapping potentials, and the energy exchange is mediated via the long-range ion-molecule polarization potential and the additional ion trap modulation allowing for an efficient exchange of phonons between the two particles. We have considered two distinct cooling schemes. In the first method the laser cooling and the energy exchange pulses are separated, while in the second scheme these two processes happen at the same time. In both cases the resulting cooling rate is determined by the slower of these two processes, which typically is the energy exchange process stimulated by the parametric resonance effect. For an example system of a Ca$^+$ ion interacting with a KRb polar molecule, the cooling rate of the ion can be as large as $\Gamma_i = 100$KHz \cite{Schulz2008}, while the maximal coupling frequency $\omega_c^{max}$ determined in Section~\ref{Sec:Mod} is limited to $2 \pi \times 27$KHz for trapping frequencies $\omega_i = 2\pi \times 1$MHz and $\omega_m=2 \pi \times 100$kHz.
Assuming a slightly smaller value for the coupling $\omega_c=2\pi \times 10$KHz, for which the linear expansion of the potential is applicable, one gets the maximal cooling rate $\Gamma = 2 \Omega_c \approx 6$kHz for a moderate value of the modulation amplitude $A =0.3$. This can be further increased up to $\Gamma \approx  2 \omega_c  = 20$kHz for a modulation amplitude $A \approx 1$, at the expense of a higher energy of the molecule at the end of the cooling process.

\begin{acknowledgments}
We acknowledge support from the Polish Government Research Grant for years 2007-2010, from the DFG (SFB/TRR 21), from the EC (grant 247687, AQUTE), and the SFB FOQUS.
\end{acknowledgments}

\begin{appendix}

\section{Derivation of the effective Hamiltonian for trapped ion and molecule}
\label{App:Micro}

In this Appendix we present a microscopic derivation of the effective Hamiltonian \eqref{H0} in an adiabatic approximation and we discuss its validity. We consider an ion of mass $m_i$ with positive charge $+e$. For simplicity we do not take into account the internal structure of the ion, which has a different electronic structure than the molecule, and is not affected by the optical trapping potential. We assume that the diatomic molecule is in its ground vibrational state ($v =0$), and we treat it as a rigid rotor, neglecting the couplings to excited vibrational states. The molecule consists of two nuclei denoted as $A$ and $B$, with charges $Z_A e$ and $Z_B e$, and masses $m_A$ and $m_B$, respectively. For simplicity we carry out our derivation in the nonrelativistic approximation, neglecting the effects of fine and hyperfine level splitting. The total Hamiltonian can be written as
\begin{equation}
H = H_\mrm{kin} + H_\mrm{el} + H_\mrm{las}+ H_\mrm{rf} + H_\mrm{int}.
\end{equation}
The kinetic energy of the nuclei is
\begin{equation}
H_\mrm{kin} = \frac{\mbf{p}_i^2}{2 m_i} + \frac{\mbf{p}_m^2}{2 M} + B \mbf{J}^2,
\end{equation}
where the labels $i$ and $m$ denote, respectively, the ion and molecule COM degrees of freedom, $M = m_A + m_B$, $B J^2$ is the kinetic energy of a rigid rotor, while $B$ denotes the rotational constant of the molecule in the vibrational ground state ($v =0$). In our derivation all the unprimed quantities refer to the space-fixed frame, whereas the primed ones refer to the rotating, molecule-fixed frame.

The term $H_\mrm{el}$ is the electronic Hamiltonian, including the kinetic energy and Coulomb interactions
\begin{equation}
H_\mrm{el} = \sum_{j} \left[\frac{\mbf{p}_j^2}{2 m_\mrm{e}} - \frac{Z_A e^2}{|\mbf{r}_j - \mbf{r}_A|} - \frac{Z_B e^2}{|\mbf{r}_j - \mbf{r}_B|} \right] + \sum_{j>j'} \frac{e^2}{|\mbf{r}_j - \mbf{r}_{j'}|},
\end{equation}
with the label $j$ enumerating the electrons of the molecule.
%Here, for simplicity, we omit the contributions of the core regions for both the atom and ion. In this way our model refers in fact to H$_2^+$ molecule. For alkali atoms and alkaline-earth ions one should treat the complete structure of core regions, however, our approach can be readily generalized to this more complicated case.

The term $H_\mrm{las}$ describes the interaction of the molecule with a far-off-resonance laser light creating an optical potential. In the electric dipole representation
\begin{equation}
H_\mrm{las} = - \mbf{d} \mbf{E}_\mrm{las}(\mbf{r}_m,t),
\end{equation}
where $\mbf{d} = e (Z_A \mbf{r}_A + Z_B \mbf{r}_B - \sum_j \mbf{r}_j)$ is the dipole moment of the molecule. We have applied the long-wavelength approximation, neglecting changes of the electric field $\mbf{E}_\mrm{las}$ on the scale of the molecule. The molecule can be optically trapped, for instance, by a pair of counter-propagating laser beams with circular polarization \cite{Micheli}, creating a standing wave
\begin{align}
\mbf{E}_\mrm{las}(\mbf{r},t) & = \cos (k_L x+ \phi_x)^2 \left( E_0 \mbf{e}_+ e^{i \omega_L t}  + c.c. \right) \nonumber \\
& + \sin(k^\prime_L x+ \phi_x)^2 \left( E^{\prime}_0 \mbf{e}_+ e^{i \omega^\prime_L t}  + c.c. \right) \nonumber \\
& + ( x \rightarrow y \rightarrow z).
\end{align}
Here, the abbreviation $(x \rightarrow y \rightarrow z)$ denotes the rest of the terms obtained by cyclic permutations of the $x$, $y$ and $z$ coordinates. For each Cartesian direction $\{\mbf{e}_{x},\mbf{e}_{y},\mbf{e}_{z}\}$ we include two pairs of laser beams with amplitudes $E_0$, $E^{\prime}_0$, wave vectors $k_L$, $k^{\prime}_L$, and frequencies $\omega_L = k_L c$, $\omega^\prime_L = k^{\prime}_L c$. This allows to cancel the influence of the tensor shifts, inducing a position dependent splitting of the rotational levels \cite{Micheli}. For simplicity we have assumed the same amplitude, wave vectors, and frequencies for all three directions of the laser beams. The versors of the spherical basis are denoted by $\{\mbf{e}_{-},\mbf{e}_{0},\mbf{e}_{+}\}$, where $\mbf{e}_0 = \mbf{e}_z$, $\mbf{e}_{\pm} = \mp (\mbf{e}_x \pm i \mbf{e}_y)/\sqrt{2}$, and  $\phi_{x,y,z}$ are the phase factors.

The Hamiltonian of the radio-frequency (RF) field $H_\mrm{rf}$ creating the trapping potential for the ion reads
\begin{equation}
\label{Hrf}
H_\mrm{rf} = e \Phi_\mrm{rf}(\mbf{r}_i,t) - \mbf{d} \mbf{E}_\mrm{rf}(\mbf{r}_m,t),
\end{equation}
where $\Phi_\mrm{rf}(\mbf{r},t)$ is the time-dependent electric field of the RF trap:
\begin{align}
\Phi_\mrm{rf}(\mbf{r},t) = & \frac{1}{2} (u_x x^2 + u_y y^2 + u_z z^2) \cos \omega_\mrm{rf} t.
\end{align}
Here, $\omega_\mrm{rf}$ is the frequency of the time-dependent electric potential, and $u_k$ ($k=x,y,z$) are amplitudes depending on the trap geometry \cite{Leibfried}. The electric field at every instant of time fulfills the Laplace equation $\Delta \Phi =0$. Hence, the coefficients $u_k$, $v_k$ are subject to the following condition:
$u_x+u_y+u_z =0$. For simplicity we have assumed that the RF trap is created only by the time dependent component, which allows to avoid the effects of molecule polarization in the electric field of the RF trap. In the second term of \eqref{Hrf} describing the interaction of the molecule with the RF field, we have assumed the size of the molecule to be much smaller than the characteristic length scale of the electric field $\mbf{E}_\mrm{rf}(\mbf{r},t) = - \mbf{\nabla} \Phi_\mrm{rf}(\mbf{r},t)$.

The last term in the Hamiltonian describes the interaction of the molecule with the ion
\begin{equation}
H_\mrm{int} = -\mbf{d} \mbf{E}_i (\mbf{r}_m-\mbf{r}_i),
\end{equation}
where $\mbf{E}_i(\mbf{r}) = e \mbf{r}/r^3$. We have only included the lowest order term in the multipole expansion, assuming the ion-molecule distance to be much larger than the size of the molecule. This breaks down at shorter distances, when the ion starts to distinguish separate components of the molecule, however, the details of the short-range physics of the ion-molecule interactions are not relevant for our analysis.

Below we indicate the basic steps of the derivation.

{\em Expansion in the basis of Born-Oppenheimer wave functions for the electron motion}. We start from generating a complete set of electronic wave functions, parameterized by the positions of the molecule COM, $\mbf{r}_m$ and by the orientation $(\theta,\phi)$
\begin{equation}
H_\mrm{el} \Phi_n(\{\mbf{r}_j\}|\mbf{r}_m,\theta,\phi) = E_n \Phi_n(\{\mbf{r}_j\}|\mbf{r}_m,\theta,\phi),
\end{equation}
where $\{\mbf{r}_j\}$ denotes the set of electronic coordinates. In this way the total wave function can be expanded in the basis of Born-Oppenheimer electronic wave functions
\begin{equation}
\begin{split}
\label{Exp}
\Psi(\{\mbf{r}_j\},\mbf{r}_i,\mbf{r}_m,\theta,\phi,t) = \sum_{n} & c_n(\mbf{r}_i,\mbf{r}_m,\theta,\phi,t) \\ & \Phi_n(\{\mbf{r}_j\}|\mbf{r}_m,\theta,\phi).
\end{split}
\end{equation}
Since the basis is complete, the expansion of the wave function does not involve any approximations.

{\em Born-Oppenheimer approximation}.
%In the expansion \eqref{Exp} we keep only two modes coupled by the laser light creating the optical lattice: the electronic ground state $\Phi_g(\mbf{x}_e|\mbf{x}_1,\mbf{x}_2)$ and the electronic excited state $\Phi_e(\mbf{x}_e|\mbf{x}_1,\mbf{x}_2)$.
We use the Born-Oppenheimer method, treating the
electron motion in the adiabatic approximation. This assumes that
the time scale of the electron dynamics is much faster than the
dynamics of the molecule's nuclei, which is typically
fulfilled since the electron is much lighter than the other two
particles. Substituting expansion \eqref{Exp} into the time-dependent Schr\"odinger equation, we obtain a set of coupled differential equations for the expansion coefficients $c_n(\mbf{r}_i,\mbf{r}_m,\theta,\phi,t)$:

\begin{widetext}
\begin{align}
\label{dcg}
i \hbar \frac{\partial c_n}{\partial t} & =  \Big[
H_\mrm{kin} + e \Phi_\mrm{rf}(\mbf{r}_i,t) + E_n \Big] c_n -
\sum_m \langle\Phi_n| \mbf{d} \mbf{E}_\mrm{las}(\mbf{r}_m,t)+ \mbf{d} \mbf{E}_\mrm{rf}(\mbf{r}_m,t) + \mbf{d} \mbf{E}_i (\mbf{r}_m-\mbf{r}_i)| \Phi_m \rangle c_m.
\end{align}
\end{widetext}
%where $\mbf{d}_\mrm{eg}(\mbf{x}_1,\mbf{x}_2) = \langle \Phi_e|\mbf{d}|\Phi_g \rangle$ is the dipole matrix element between the ground and excited electronic states, which in general depends on the position of atom and ion.
{\em Adiabatic elimination of the excited electronic states: derivation of the optical trap potential}.
For a far-detuned laser, the excited states are only weakly populated and can be adiabatically eliminated. In the spirit of the Born-Oppenheimer approximation we can assume that the transitions between ground and excited electronic states, due to the laser light, occur on a time scale much shorter than the motion of the molecule and the ion, and the dynamics of their COM motion can be decoupled from the internal dynamics. The result of the adiabatic elimination is most conveniently expressed in terms of the dynamic polarizabilities $\alpha_\|(\omega)$ and $\alpha_\perp(\omega)$ in the directions parallel and perpendicular to the internuclear axis \cite{Friedrich}. The calculation can be performed in a similar manner as described in Ref.~\cite{Micheli}, with the only difference that here we consider a three-dimensional optical lattice. As a result we obtain the following trapping potential for a molecule:
\begin{equation}
V_\mrm{opt}(\mbf{r}) = - V_0 + \tst{\frac12} M \omega_m^2 (\mbf{r}-\bar{\mbf{r}})^2,
\end{equation}
where $V_0$ is the position independent component of the AC Stark shift $V_0 = |E_0|^2 \left[\frac{2}{3} \alpha_\|(\omega_L) +\frac13 \alpha_\perp(\omega_L)\right]$, and the trapping frequency is
given by
\begin{align}
\omega_m^2 = \frac2M \bigg\{ & |E_0 k_L|^2 \left[ \tst{\frac23} \alpha_\|(\omega_L) + \tst{\frac13} \alpha_\perp(\omega_L) \right] \nonumber \\
& - |E^\prime_0 k^\prime_L|^2 \left[ \tst{\frac23} \alpha_\|(\omega^\prime_L) + \tst{\frac13} \alpha_\perp(\omega^\prime_L) \right] \bigg\}.
\end{align}
The second laser with frequency $\omega_L^\prime$ is blue detuned from the resonance in order to make $\alpha_\|(\omega)$ and $\alpha_\perp(\omega)$ both negative. With this assumption $\omega_m^2$ is positive, and one can additionally adjust the parameters of the second laser to fulfill
\begin{multline}
|E_0 k_L|^2 \left[ \tst{\frac23} \alpha_\|(\omega_L) - \tst{\frac23} \alpha_\perp(\omega_L) \right] \\
- |E^\prime_0 k^\prime_L|^2 \left[ \tst{\frac23} \alpha_\|(\omega^\prime_L) - \tst{\frac23} \alpha_\perp(\omega^\prime_L) \right] = 0,
\end{multline}
which allows to cancel position-dependent terms sensitive to the orientation of the molecule. We note that in the case of a three-dimensional lattice created by laser beams of identical intensity, the AC Stark shift in the center of the lattice well ($\mbf{r} =0$) does not depend on the molecule orientation, in contrast to a one-dimensional optical lattice \cite{Micheli}.

{\em Time averaging over fast oscillations of the RF field: derivation of the Paul trapping potential}. We replace the time-dependent RF field by an effective adiabatic potential for the ion, neglecting its fast micromotion on the time scale of $\omega_\mrm{rf}$ \cite{Leibfried}. Following the standard approach described in \cite{Leibfried}, we obtain
\begin{equation}
V_\mrm{rf}(\mbf{r}) = \frac{1}{2} m_i \left(\omega_x x^2 +\omega_y y^2 + \omega_z z^2 \right),
\end{equation}
where $\omega_k = (q_k^2/2)^{1/2} \omega_\mrm{rf} /2$, with $q_k = 2 e u_k/(m_2 \omega_\mrm{rf}^2)$  for $k=x,y,z$. We can estimate the influence of the RF field on the molecule by considering the corresponding AC Stark shift $\Omega^2/\Delta$, resulting from the transitions $J=0 \rightarrow J=1$. For a typical molecule ($d \sim 1D$) and for electric fields down to distances of the order of few tens of harmonic oscillator lengths from the trap center, the Rabi frequency $\Omega \ll \omega_\mrm{rf}$. The rotational splitting given by $B$ is typically much larger than $\omega_\mrm{rf} \lesssim 10$MHz, therefore the detuning is of the order of $\Delta \sim B \sim 1$GHz and the resulting AC Stark shift of the molecule in the RF field can be neglected in comparison to the frequency of the optical potential.

Now, including the adiabatic elimination and time-averaging of the RF potential, we obtain the following equation for the wave function $c_g(\mbf{r}_i,\mbf{r}_m,\theta,\phi,t)$ corresponding to the electronic ground state:
\begin{align}
i \hbar \frac{\partial c_g}{\partial t}  =  \bigg[ &
H_\mrm{kin} + V_\mrm{rf}(\mbf{r}_i)+V_\mrm{opt}(\mbf{r}_m) + E_g \nonumber  \\
& - \mbf{d}_{gg} \mbf{E}_i (\mbf{r}_m-\mbf{r}_i) \bigg] c_g(\mbf{r}_i,\mbf{r}_m,\theta,\phi,t).
\end{align}
Here, $\mbf{d}_{gg} = \langle \Phi_g | \mbf{d} | \Phi_g \rangle$ is the dipole moment of the molecule calculated in the electronic ground state. It depends on the orientation of the internuclear axis, and its coordinates in the space-fixed spherical basis are given by $(d_{gg})_q = \mbf{d}_{gg} \cdot \mbf{q} = d_{gg} C_q^{(1)}(\theta,\phi)$ for $q=0,\pm 1$.

{\em Adiabatic elimination of the molecule rotational degrees of freedom.} In analogy to the Born-Oppenheimer approximation for the electronic wave functions, we can perform an adiabatic approximation with respect to the ionic and molecular translational degrees of freedom. This can be done since the molecular rotations occur on a timescale $2 \pi/B$, which is much shorter that the time scale of the translational motion, given by $2 \pi/\omega_i$ and $2 \pi/\omega_m$ for the ion and the molecule, respectively. The total wave function can be expanded in the basis of wave functions $\Lambda_n(\theta,\phi|\mbf{r}_i,\mbf{r}_m)$ of the rotational motion, parameterized by the positions of the ion and of the molecule
\begin{equation}
c_g(\mbf{r}_i,\mbf{r}_m,\theta,\phi,t) = \sum_{n} \psi_n(\mbf{r}_i,\mbf{r}_m,t) \Lambda_n(\theta,\phi|\mbf{r}_i,\mbf{r}_m),
\end{equation}
which fulfill
\begin{multline}
\label{BOrot}
\left[B \mbf{J}^2 - \mbf{d}_{gg} \mbf{E}_i (\mbf{r}_m-\mbf{r}_i) \right] \Lambda_n(\theta,\phi|\mbf{r}_i,\mbf{r}_m) = \\
=E_n(\mbf{r}_m-\mbf{r}_i) \Lambda_n(\theta,\phi|\mbf{r}_i,\mbf{r}_m).
\end{multline}
The solutions of \eqref{BOrot} can be found by expanding in the basis of spherical harmonics $Y_{JM}(\theta,\phi)$
\begin{equation}
\Lambda_n(\theta,\phi|\mbf{r}_i,\mbf{r}_m) = \sum_{JM} d_{JM} Y_{JM}(\theta,\phi),
\end{equation}
where the $z$ axis is chosen along $\mbf{r} = \mbf{r}_m - \mbf{r}_i$. Introducing a dimensionless parameter $\eta = d_{gg} E_i(|\mbf{r}|)/B$ characterizing the coupling between angular momenta $J$ and $J \pm 1$, we observe that typically $\eta \ll 1$ at ion-molecule distances corresponding to our cooling scheme. Therefore the eigenstates $\Lambda_n$ are dominated by  $|J,M\rangle$ with a small admixture of $|J\pm1,M\rangle$ proportional to $\eta$. In this case we can solve for $E_n$ and $\Lambda_n$, taking into account only $|J,M\rangle$, $|J\pm1,M\rangle$ states, and neglecting couplings to other states. This yields
\begin{equation}
E_{JM} = B\left[J(J+1) + \frac{\eta^2}{2} \frac{J(J+1)-3 M^2}{J(J+1)(2J-1)(2J+3)} \right],
\end{equation}
which is the sum of the unperturbed eigenvalue plus a small correction. For a molecule in the rotational ground state the eigenenergy is $E_{00} = - B \eta^2/6$, which leads to an attractive potential between the ion and the molecule
\begin{equation}
V_\mrm{int}(r) =  - \frac{(d_{gg} e)^2}{6 B} \frac{1}{r^4}.
\end{equation}

\section{diagonalization of the coupled oscillator Hamiltonian}
\label{App:Transf}

The Hamiltonian \eqref{Hz} can be diagonalized with the help of the following transformation:
\begin{align}
a^\dagger + a &= c_\alpha (\alpha^\dagger + \alpha)  + c_\beta (\beta^\dagger +\beta) ,\\
b^\dagger + b &= d_\alpha (\alpha^\dagger + \alpha)  + d_\beta (\beta^\dagger +\beta) ,\\
a^\dagger - a &= W \left[d_\beta (\alpha^\dagger - \alpha) - d_\alpha (\beta^\dagger -\beta) \right],\\
b^\dagger - b &= W \left[c_\alpha(\beta^\dagger -\beta) - c_\beta (\alpha^\dagger - \alpha) \right],
\end{align}
where
\begin{align}
c_\alpha &= \sqrt{\frac{\omega_i (\omega_2^2-\omega_i^2)}{\Omega^2 \omega_1}},
&c_\beta = \sqrt{\frac{\omega_i (\omega_i^2-\omega_1^2)}{\Omega^2 \omega_2}}, \\
d_\alpha &= -\sqrt{\frac{\omega_m (\omega_i^2-\omega_1^2)}{\Omega^2 \omega_1}},
&d_\beta = \sqrt{\frac{\omega_m (\omega_2^2-\omega_i^2)}{\Omega^2 \omega_2}},
\end{align}
$\omega_1$ and $\omega_2$ are the frequencies of the two eigenmodes $\omega_{1,2}^2  = \frac{1}{2}\left[\omega_i^2 + \omega_m^2 \mp \Omega^2 \right]$
with the upper (lower) sign referring to the frequency $\omega_1$ ($\omega_2$), respectively, and
\begin{align}
\Omega^2 & =\sqrt{(\omega_i^2 - \omega_m^2)^2 + 16 \omega_c^2 \omega_m \omega_i},\\
W & = \sqrt{\frac{\omega_1 \omega_2}{\omega_i \omega_m}}.
\end{align}

\section{periodic modulation of the trapping frequency for an ion confined in an RF trap}
\label{App:Stability}

\begin{figure}
   \includegraphics[width=8.6cm,clip]{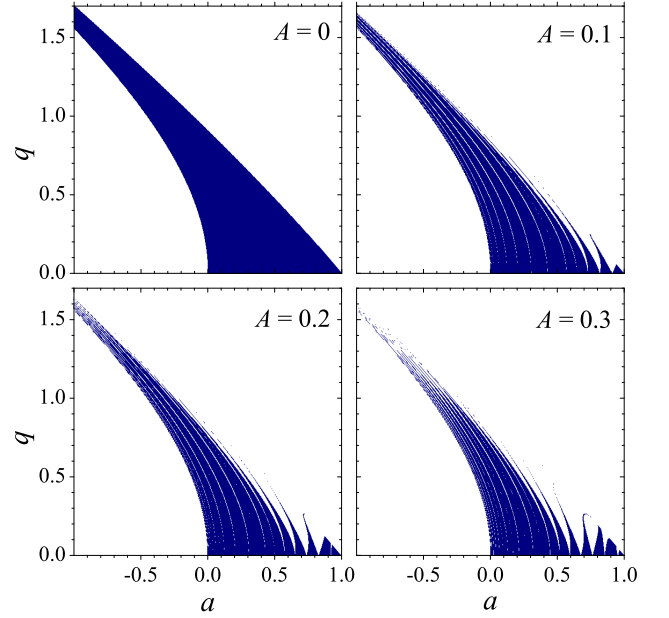}
	 \caption{
	 Stability diagram of Eq.~(\ref{HillsEq}) for $\omega_\mrm{f}/\omega_\mrm{rf}=0.05$,
	 and $A=0$, $0.1$, $0.2$ and $0.3$. The amplitude
	 $W$ of the component varying with the frequency $\omega_\mrm{f}$
	 is determined from the parameters $a$, $q$ and $A$ (see text for details).
	 \label{Fig:Stab}
   }
\end{figure}

For an ion stored in RF trap, a periodic modulation of the trapping frequency can be realized by adding to the electric
field an additional component, which varies with a frequency $\omega_\mrm{f}$.
To be more precise, the electric potential felt by the ion
in the center of the trap should be of the form $\Phi(\mbf{r})=
(U + V \cos \omega_\mrm{rf} t + W \cos \omega_\mrm{f} t) (\alpha_x x^2
+ \alpha_y y^2 + \alpha_z z^2)/(2 r_0^2)$, where $\omega_\mrm{rf}$ is the frequency of the RF field, $r_0$ is the
characteristic distance describing the size of the RF trap, and
$\alpha_x$, $\alpha_y$, $\alpha_z$ are geometrical factors depending on the spatial configuration of the trap \cite{Leibfried}. At every instant in time $\Phi(\mbf{r})$ has to fulfill the Laplace equation: $\Delta \Phi =0$, hence $\alpha_x+ \alpha_y+\alpha_z=0$. Classically, the motion of the ion in the direction $k$ ($k = x,y,z$) is governed by
\begin{equation}
\label{HillsEq}
m_i \frac{d^2 x_k}{d t^2} + \frac{\alpha_k e}{r_0^2}(U + V \cos \omega_\mrm{rf} t + W \cos \omega_\mrm{f} t)  x =0
\end{equation}
For $W=0$, Eq. (\ref{HillsEq}) reduces to the well known
Mathieu equation. Its lowest stability region is presented in the first panel of Fig.~\ref{Fig:Stab} as a function of the parameters
$a =4 \alpha_k e U/(m_i \omega_\mrm{rf}^2 r_0^2)$ and $q=2 \alpha_k e V/(m_i \omega_\mrm{rf}^2 r_0^2)$.
The motion is stable in the gray-shaded regions of the
$a$-$q$ plane. Inclusion of a field varying with frequency $\omega_\mrm{f}$ modifies this stability diagram,
which can be seen from Fig.~\ref{Fig:Stab} presenting the regions of stability calculated for
$\omega_\mrm{f}/\omega_\mrm{rf}=0.05$ and $A=0.1$, $0.2$, and $0.3$.
At each point on the $a$-$q$ plane, the amplitude $W$ is adjusted in such a way that the motion of the ion, averaged
over the fast micromotion on the time-scale of $\omega_\mrm{rf}$, can be described effectively as a motion in the harmonic
trap with the modulated frequency $\omega_i^2(t) = \omega_i^2 (1 + 2 A \cos(\omega_f t))$. Finally, the parameters of the
RF trap have to be chosen in such a way that the motion of the ion is stable simultaneously for all three spatial directions.
%%%%%%%%%%%%%%%%%% Figure 13 %%%%%%%%%%%%%%%%%%%%%%
%%%%%%%%%%%%%%%%%%%%%%%%%%%%%%%%%%%%%%%%%%%%%%%%%%%%%

\section{Adiabatic elimination}
\label{App:AdiabElim}

In this section we perform adiabatic elimination of the internal dynamics. The basic steps of this derivation are similar
to the treatment of laser cooling of a single ion (cf. for instance Ref.~\cite{Cirac}). First we expand the ME
\eqref{ME} in powers of the Lamb-Dicke parameter $\eta = k \sqrt{\hbar/(2 m_i \omega_i)} \ll 1$,
retaining terms up to the second order in $\eta$:
\begin{equation}
\label{MEexp}
\dot{\rho} = \left( {\cal L}_{0I} + {\cal L}_{0E} + {\cal L}_1 + {\cal L}_2 \right) \rho(t),
\end{equation}
where
\begin{align}
\label{L0I}
{\cal L}_\mrm{0I} \rho & =   \frac{1}{i \hbar}
\left[ H_\mrm{las}^{(0)} + H_\mrm{int},\rho \right] \nonumber \\
& \quad \, + \frac{\gamma}{2} \left[ 2 \sigma_{-} \rho \sigma_{+} - \sigma_{+} \sigma_{-} \rho -
\rho \sigma_{+} \sigma_{-} \right], \\
\label{L0E}
{\cal L}_\mrm{0E} \rho & =  \frac{1}{i \hbar} \left[ H, \rho \right], \\
{\cal L}_{1} \rho & = \frac{1}{i \hbar} \left[ H_\mrm{las}^{(1)}, \rho \right], \\
{\cal L}_{2} \rho & = \frac{1}{i \hbar} \left[ H_\mrm{las}^{(2)},
\rho \right]+\frac{\gamma}{5} k^2 \sigma_{-}
\left(2 z_i \rho z_i - z_i^2 \rho - \rho z_i^2\right) \sigma_{+}.
\end{align}
Here $H_\mrm{las}^{(0)}$, $H_\mrm{las}^{(1)}$, and $H_\mrm{las}^{(2)}$ are the terms of expansion of $H_\mrm{las}$ in
the powers of $\eta$:
\begin{align}
H_\mrm{las}^{(0)} & = \frac{\hbar \Omega}{2} (\sigma_{-} + \sigma_{+}), \\
H_\mrm{las}^{(1)} & = \frac{\hbar \Omega}{2} (i k z_i \sigma_{+} + h.c.), \\
H_\mrm{las}^{(2)} & = - \frac{\hbar \Omega}{4} (k^2 z_i^2 \sigma_{+} + h.c.).
\end{align}
To lowest order in $\eta$, the dynamics is described by terms ${\cal L}_\mrm{0I}$ and
${\cal L}_\mrm{0E}$, and the evolution
of internal and external degrees of freedom is decoupled. The
adiabatic elimination can be done provided that the internal dynamics is faster than the lowest-order
coupling between the internal and external degrees of freedom, which is given by ${\cal L}_{1}$.
Since the time-scale of the internal dynamics is given by $\Omega$ and $\gamma$ (cf. Appendix~\ref{App:TTC}),
the condition for the adiabatic elimination can be written as $\eta \Omega \ll \gamma,\Omega$.

The adiabatic elimination is done with the help of the projection operators technique \cite{Gardiner}.
We introduce the
operator ${\cal P} \rho = \rho_{s} \mrm{Tr}_I \{\rho\}$, where the trace is taken over the internal degrees of freedom,
and $\rho_{s}$ denotes the stationary density matrix of the internal states: ${\cal L}_{0I} \rho_{s} \equiv 0$. First
we project the ME \eqref{MEexp} using ${\cal P}$ and the complementary operator ${\cal Q}\equiv 1 - {\cal P}$. Then
applying the identities ${\cal P} {\cal L}_\mrm{0E} = {\cal L}_\mrm{0E} {\cal P} $,
${\cal P} {\cal L}_\mrm{0I} = {\cal L}_\mrm{0I} {\cal P}=0$ we obtain the following set of differential equations
for the functions $u(t) = {\cal P} \rho(t)$ and $w(t) = {\cal Q} \rho(t)$
\begin{align}
\label{udot}
\dot{u}(t) & = \left({\cal L}_\mrm{0E} + {\cal P} ({\cal L}_{1} +{\cal L}_{2}) \right) u +
{\cal P} ({\cal L}_{1} +{\cal L}_{2}) w\\
\label{wdot}
\dot{w}(t) & = {\cal Q} ({\cal L}_{1} +{\cal L}_{2}) w +
\left({\cal L}_\mrm{0E} + {\cal L}_\mrm{0I} + {\cal Q} ({\cal L}_{1} +{\cal L}_{2}) \right) u,
\end{align}
We solve \eqref{udot}-\eqref{wdot} for $u(t)$ using the Laplace transformation. This leads us to
\begin{equation}
\label{uelim}
\begin{split}
\dot{u}(t) & = \left({\cal L}_\mrm{0E} + {\cal P} ({\cal L}_{1} +{\cal L}_{2}) \right) u(t) \\
& \quad \; + {\cal P} ({\cal L}_{1} +{\cal L}_{2}) \!\!\int_0^\infty \!\!\! \mrm{d}\tau \bigg[
e^{\tau ({\cal L}_\mrm{0E} + {\cal L}_\mrm{0I} + {\cal Q} ({\cal L}_{1} +{\cal L}_{2}))} \\
& \qquad \qquad \qquad \qquad \qquad \; \times {\cal Q} ({\cal L}_{1} +{\cal L}_{2}) u(t-\tau) \bigg].
\end{split}
\end{equation}
In the next step we apply the Markov approximation to the function under integral:
$u(t-\tau) \approx u(t)$, and we neglect in \eqref{uelim}  all the terms of order higher than $\eta^2$. In this
way we arrive at
\begin{equation}
\label{uapprox}
\begin{split}
\dot{u}(t) & = \left({\cal L}_\mrm{0E} + {\cal P} ({\cal L}_{1} +{\cal L}_{2}) \right) u(t) \\
& \quad \; + {\cal P} {\cal L}_{1} \!\!\int_0^\infty \!\!\! \mrm{d}\tau \bigg[
e^{\tau ({\cal L}_\mrm{0E} + {\cal L}_\mrm{0I})} {\cal Q} {\cal L}_{1} u(t) \bigg],
\end{split}
\end{equation}
Finally we trace over the internal degrees of freedom obtaining the ME for the reduced density operator
$\rho_e = \mrm{Tr}_I \{\rho\}$. A calculation of the trace of the first term on r.h.s. of \eqref{uapprox} gives
\begin{align}
\mrm{Tr}_I \{{\cal L}_\mrm{0E} u(t) \} & =  {\cal L}_\mrm{0E} \rho_e \\
\mrm{Tr}_I \{{\cal P} {\cal L}_1 u(t) \} & = i \frac{k \Omega}{2} \langle \sigma_y \rangle_{s} [z_i,\rho_e],\\
\mrm{Tr}_I \{{\cal P} {\cal L}_2 u(t) \} & = i \frac{k^2 \Omega}{4} \langle \sigma_x \rangle_{s} [z_i^2,\rho_e]
\nonumber \\
& \quad \, + \frac{\gamma}{5} k^2 \langle \sigma_{+} \sigma_{-} \rangle_{s}
\left(2 z_i \rho_e z_i - z_i^2 \rho_e - \rho_e z_i^2 \right),
\end{align}
where $\langle . \rangle_{s} \equiv \textrm{Tr}_I \{ . \rho_{s} \}$ denotes an average taken in the stationary state of
the density matrix $\rho_{s}$ for the internal degrees of freedom. The calculation of the trace of the second term in Eq.~\eqref{uapprox} is more involved, and as a result we obtain
\begin{multline}
\label{Tr2}
\mrm{Tr}_I \{ {\cal P} {\cal L}_{1} \!\!\int_0^\infty \!\!\! \mrm{d}\tau
e^{\tau ({\cal L}_\mrm{0E} + {\cal L}_\mrm{0I})} {\cal Q} {\cal L}_{1} u(t) \} = \\
= \frac{k^2 \Omega^2}{4} \! \! \int_0^{\infty} \! \! \! \mrm{d}\tau \,  \bigg[
\left( z_i(\tau) \rho_e z_i - z_i z_i(\tau) \rho_e \right) \\
\times \Big(\langle \sigma_y(\tau) \sigma_y(0) \rangle_{s}
- \langle \sigma_y \rangle_{s}^2 \Big) + h.c. \bigg].
\end{multline}
In the derivation of \eqref{Tr2} we have applied the following approximation
$e^{\tau {\cal L}_\mrm{0E}} (z_i \rho_e(t)) = (e^{\tau {\cal L}_\mrm{0E}} z_i) (e^{\tau {\cal L}_\mrm{0E}} \rho_e(t))
\approx (e^{\tau {\cal L}_\mrm{0E}} z_i) \rho_e(t)$, which is compatible with the Markov approximation used in the
above equations. The symbol $z_i(\tau)$ denotes the operator evolved according to ${\cal L}_\mrm{0E}$:
\begin{equation}
z_i(\tau)  = e^{\tau {\cal L}_\mrm{0E}} z_i = e^{- i t H /\hbar} z_i  e^{i t H/\hbar},
\end{equation}
while
\begin{equation}
\langle \sigma_y(\tau) \sigma_y(0) \rangle_{s} \equiv
\textrm{Tr}_I \left\{ \sigma_y (e^{\tau {\cal L}_{0I}} \sigma_y \rho_{ss}) \right\}
\end{equation}
is a two-time correlation function, discussed in more details in the next section.

\section{Dynamics of the internal degrees of freedom}
\label{App:TTC}

In this section we analyze the dynamics of the internal degrees of freedom, which to lowest order in the Lamb-Dicke parameter is governed by
\begin{equation}
\label{MEint}
\dot{\rho}(t) = {\cal L}_\mrm{0I} \rho,
\end{equation}
with ${\cal L}_\mrm{0I}$ defined in \eqref{L0I}. We start by writing Bloch equations for the averages of the
ion internal operators: $\langle \sigma_x \rangle$, $\langle \sigma_y \rangle$, and $\langle \sigma_z \rangle$,
which follow from the ME \eqref{MEint}
\begin{align}
\label{dsx}
\frac{d}{dt} \langle \sigma_x \rangle & = \Delta \langle \sigma_y \rangle - \frac{\gamma}{2}
\langle \sigma_x \rangle, \\
\frac{d}{dt} \langle \sigma_y \rangle & = - \Omega \langle \sigma_z \rangle - \Delta
\langle \sigma_x \rangle - \frac{\gamma}{2} \langle \sigma_y \rangle, \\
\label{dsz}
\frac{d}{dt} \langle \sigma_z \rangle & = \Omega \langle \sigma_y \rangle - \gamma
\left(1+ \langle \sigma_z \rangle \right).
\end{align}
From Eqs.~\eqref{dsx}-\eqref{dsz} one can easily calculate the average values of the internal operators in the stationary state
\begin{align}
\label{avgsx}
\langle \sigma_x \rangle_s & = - \frac{ \Omega \Delta}{\delta^2}, \\
\langle \sigma_y \rangle_s & = \frac{ \Omega \gamma}{2 \delta^2}, \\
\label{avgsz}
\langle \sigma_z \rangle_s & = - \frac{ \gamma^2 + 4 \Delta^2}{4 \delta^2},
\end{align}
where $\delta^2 = \frac12 \Omega^2 + \frac14 \gamma^2 + \Delta^2$. Now we utilize the quantum regression theorem (see for instance Ref. \cite{Carmichael}), and with the help of Eqs.~\eqref{dsx}-\eqref{dsz} we write the equations of motion for the two-time
correlation functions $s_k(\tau) \equiv \langle \sigma_y(0) \sigma_k(\tau) \rangle$, where $k=x,y,z$,
\begin{align}
\dot{s}_x(\tau) & = \Delta s_y - \frac{\gamma}{2} s_x, \\
\dot{s}_y(\tau) & = - \Omega s_z - \Delta s_x - \frac{\gamma}{2} s_y, \\
\dot{s}_z(\tau) & = \Omega \langle \sigma_y \rangle - \gamma s_z - \gamma \langle \sigma_y \rangle.
\end{align}
Applying the Laplace transformation we convert the set of differential equations into a set of algebraic equations and
solving for $\tilde{s}_y(s) = \int_0^\infty \mrm{d}\tau \, e^{-s \tau} s_y(\tau)$ we find
\begin{widetext}
\begin{equation}
\tilde{s}_y(s) = \frac{(s+\gamma)(s+\frac{\gamma}{2})s_y(0)-(s+\gamma)\Delta s_x(0) -
\Omega (s+\frac{\gamma}{2})\left(s_z(0) - \Omega \gamma^2/(2 \delta^2 s) \right)}{
(s+\gamma) \left( \Delta^2 + (s+\frac{\gamma}{2})^2\right) + \Omega^2 (s+\frac{\gamma}{2})}.
\end{equation}
\end{widetext}
The values of the correlation function $s_k(\tau)$ at $\tau =0$ can be easily expressed in terms of the
averages \eqref{avgsx}-\eqref{avgsz}: $s_x(0) = - i \langle \sigma_z \rangle_s$, $s_y(0) = 1$,
$s_z(0) = i \langle \sigma_x \rangle_s$. Now observing that
\begin{equation}
\int_{0}^\infty \!\! d\tau \,
e^{i \omega \tau} \langle \sigma_y(\tau) \sigma_y(0) \rangle_{s} = \tilde{s}_y(i \omega)^{\ast},
\end{equation}
we can calculate the function $S(\omega)$ defined in \eqref{Somega}:
\begin{equation}
S(\omega) =  \frac{\eta^2 \Omega^2}{4} \left[
\tilde{s}_y(i \omega)^{\ast} - \langle \sigma_y \rangle_{s}^2
\left( \pi \delta(\omega) + i \mrm{Pf}\left(1/\omega\right) \right) \right].
\end{equation}
where $\delta(\omega)$ is the delta function and
$\mrm{Pf}(1/\omega)$ denotes the principal value.

\section{Calculations beyond RWA}
\label{App:BeyRWA}

In this section we derive equations of motion for averages quadratic in the operators
$z_i$, $p_i$, $z_m$, $p_m$. Assuming that $\eta \Omega \ll \omega_i$, we neglect
the second and the third term in the commutator in \eqref{MEae}, describing the effects of the laser field on the
ion's trapping potential. Introducing the dimensionless variables
$x_1=\langle z_i^2 \rangle /l_i^2$, $x_2=\langle z_m^2 \rangle /l_m^2$,
$x_3=\langle p_i^2 \rangle l_i^2/\hbar^2$,
$x_4=\langle p_m^2 \rangle l_m^2/\hbar^2$, $x_5=\langle z_i p_i + p_i z_i \rangle /\hbar$,
$x_6=\langle z_m p_m + p_m z_m \rangle /\hbar$, $x_7=\langle z_m p_i \rangle l_i
/(\hbar l_m)$, $x_8=\langle z_i p_m \rangle l_m /(\hbar l_i)$, $x_9=\langle p_i p_m \rangle l_i l_m/\hbar^2$,
$x_{10}=\langle z_i z_m \rangle /(l_i l_m)$,
we obtain the following set of linear differential equations
\begin{align*}
% \label{EqsMot1}
\dot{x}_1 = & \omega_i x_5, \\
\dot{x}_2 = & \omega_m x_6, \\
\dot{x}_3 = & - \omega_i(t) x_5 + {\tst \frac{4}{5}}
\gamma \eta^2 \langle \sigma_{+} \sigma_{-} \rangle_{ss} - 4 \omega_c
x_7 \nonumber \\
& - W_1 x_5 - 2 W_2 x_3 - 2 W_4 x_9 - 2 W_3 x_7 + V_1 - U_1, \\
\dot{x}_4 = & - \omega_m x_6 - 4 \omega_c x_8, \\
\dot{x}_5  = & 2 \omega_i x_3 - 2 \omega_i(t) x_1 - 4 \omega_c x_{10}  \nonumber \\
& - W_2 x_5 - 2 W_1 x_1 - 2 W_4 x_8 - 2 W_3 x_{10} - V_2 + U_2, \\
\dot{x}_6 = & 2 \omega_m x_4 - 2 \omega_m x_2 - 4 \omega_c x_{10}, \\
\dot{x}_7 = & \omega_m x_9 - \omega_i(t) x_{10}  - 2 \omega_c x_2 \nonumber \\
& - {\tst \frac12} W_4 x_6 - W_1 x_{10} - W_2 x_7 - W_3 x_2 - V_4 + U_4, \\
\dot{x}_8 = & \omega_i x_9 - \omega_m x_{10} - 2 \omega_c x_1, \\
\dot{x}_9 = & - \omega_m x_7 - \omega_i(t) x_8 - \omega_c \left( x_5 +x_6 \right) \nonumber \\
& - {\tst \frac12} W_3 x_6 - W_1 x_8  - W_2 x_9 - W_4 x_4 + V_3 - U_3, \\
% \label{EqsMot10}
\dot{x}_{10} = & \omega_i x_7 + \omega_m x_8 .
\end{align*}
Here, $\omega_i(t) = \omega_i (1+ 2A \cos(\omega_\mrm{f} t))$,
\begin{eqnarray}
\label{Wk}
W_k &  = & \textrm{Im} \left[ \int_0^{\infty} \!\! \mrm{d}\tau \, f_k(\tau) \langle \sigma_y(\tau) \sigma_y(0)
\rangle_{s} \right], \\
V_k & = & \textrm{Re} \left[ \int_0^{\infty} \!\! \mrm{d}\tau \, f_k(\tau) \langle \sigma_y(\tau) \sigma_y(0)
\rangle_{s} \right], \\
\label{Uk}
U_k & = & \int_0^{\infty} \!\! \mrm{d}\tau \, f_k(\tau),
\end{eqnarray}
with $k=1,2,3,4$. The functions $f_k(\tau)$
are determined by the evolution of the operator $z_i$ due to ${\cal L}_\mrm{0E}$:
$z_i(\tau)/l_i \equiv e^{\tau {\cal L}_\mrm{0E}}/l_i =
f_1(\tau) z_i/l_i + f_2(\tau) p_i l_i/\hbar + f_3(\tau) z_m/l_m + f_4(\tau) p_m l_m/\hbar$.
We note that $z_i(t) = z_i(-t)_\mrm{H}$, where
$z_i(t)_\mrm{H}$ denotes the operator evolved in the Heisenberg picture. The functions $f_k(\tau)$
can be easily calculated from the differential equations derived from ${\cal L}_\mrm{0E}$ \eqref{L0E}, with
$H$ given by \eqref{H}
\begin{align}
\frac{d p_i}{d \tau}  = &  \omega_i(t) z_i + 2 \omega_c z_m, \\
\frac{d p_m}{d \tau} = &  \omega_m z_m + 2 \omega_c z_i, \\
\frac{d z_i}{d \tau} = &  - \omega_i  p_i, \\
\frac{d z_m}{d \tau} = &  - \omega_m p_m.
\end{align}

\end{appendix}

\bibliography{imc}

\begin{thebibliography}{38}
\bibitem [{foo()}]{footnote1}%
  \BibitemOpen
  \href@noop {} {}\bibinfo {note} {To find the resonance width $\delta \omega$
  we compare the value of $\langle T_3 \rangle$ after half of the oscillation
  period $T/2$, with the the prediction of formula \eqref{Tz}: $\langle
  T_3(T/2) \rangle = N (\Delta^2- (\delta \omega)^2)/(\Delta^2+ (\delta
  \omega)^2)$}\BibitemShut {NoStop}%
\end{thebibliography}

\end{document}